\documentclass[%
 reprint,
 twocolumn,
superscriptaddress,
 amsmath,amssymb,
 aps,
]{revtex4-1}

\usepackage{dcolumn}% Align table columns on decimal point
\usepackage{bm}% bold math
\usepackage{hyperref, cleveref}% add hypertext capabilities
\usepackage{amsmath,amssymb,amsfonts}
\usepackage{hyperref,url}
\usepackage{graphicx,xcolor}
\usepackage{wrapfig,tabularx}

\usepackage{color}

\begin{document}

\title{
Accurate relativistic observables from post-processing
light cone  catalogues}

\author{Chi Tian}
\email{chit@wustl.edu} 
\affiliation{Department of Physics and McDonnell Center for the Space Sciences,
Washington University, St. Louis, MO 63130, USA}

\author{Matthew F. Carney}
%\email{c.matthew@wustl.edu}
\affiliation{Department of Physics and McDonnell Center for the Space Sciences,
Washington University, St. Louis, MO 63130, USA}

\author{James B. Mertens}
%\email{jmertens@wustl.edu} 
\affiliation{Department of Physics and McDonnell Center for the Space Sciences,
Washington University, St. Louis, MO 63130, USA}

\author{Glenn Starkman}
% \email{glenn.starkman@case.edu} 
\affiliation{CERCA/ISO, Department of Physics, Case Western Reserve University, 10900 Euclid Avenue, Cleveland, OH 44106, USA}

\date{\today}

\Crefname{equation}{Eq.}{Eqs.}
\Crefname{figure}{Fig.}{Figs.}
\Crefname{tabular}{Tab.}{Tabs.}

\begin{abstract}
We introduce and study a new scheme to construct relativistic observables from post-processing light cone data. This construction is based on a novel approach, \textsc{LC-Metric}, which takes general light cone or snapshot output generated by arbitrary N-body simulations or emulations and solves the linearized Einstein equations to determine the spacetime metric on the light cone.
We find that this scheme is able to determine the metric to high precision, and subsequently generate accurate mock cosmological observations sensitive to effects such as post-Born lensing and nonlinear ISW contributions.
By comparing to conventional methods in quantifying those general relativistic effects, we show that this scheme is able to accurately construct the lensing convergence signal.
We also find the accuracy of this method in quantifying the ISW effects in the highly nonlinear regime outperforms conventional methods by an order of magnitude.
This scheme opens a new path for exploring and modeling higher-order and nonlinear general relativistic contributions to cosmological observables, including mock observations of gravitational lensing and the moving lens and Rees-Sciama effects.
\end{abstract}

\maketitle

\section{Introduction}

Observations of our Universe across cosmological distances offer us an ideal grounds for testing the behavior of long-standing physical theories, alongside new ones, ranging from signatures of new primordial phenomena \cite{1710.09465,1810.09463,1810.13424,1904.10981,1908.08953}, to models of dark energy \cite{astro-ph/0504115,Nakamura:2018oyy,Ichikawa:2005hi,Torres-Rodriguez:2007qar,Akrami:2017cir}, to general relativity (GR) itself \cite{1906.04208,1904.10033}.
On the other hand, it is often on small scales---the distances considerably smaller than our cosmological horizon on which nonlinear growth of structure occurs---that we have been able to perform our most precise measurements.
Upcoming experiments, such as CMB-S4 \cite{2006.03060} and the Vera Rubin Observatory (VRO, formerly LSST) \cite{0912.0201} will begin to provide us with a view of the Universe that encompasses both of these regimes.
Together, maps of the cosmic microwave background (CMB), of the large-scale galaxy distribution, and of other tracers of matter in the Universe, are expected to be of sufficient sensitivity and volume that we can measure or constrain a variety of the subtle effects that leave an imprint on ultra-large scales \cite{1710.02477}. 

Sophisticated, accurate theoretical predictions are essential for optimal analysis of observations at this level of precision,
and large suites of simulations are often required to make statistically meaningful predictions.
To this end, a variety of procedures designed to generate accurate mock observations without running more expensive simulations of structure formation have been proposed and explored, with varying degrees of success 
\cite{1301.0322, 1506.03737, 1509.04685, 1603.00476, 1810.07727, 1806.09499}.
These emulators can efficiently incorporate nonlinear physics in cosmological models, and can further be used to generate a variety of observations consistently in order to study e.g. cross-correlations or to test procedures such as tomographic reconstruction.
However, methods for incorporating lightcone projection effects in mock observations are underdeveloped, especially at a nonlinear level.
For unbiased extraction of cosmological parameters, especially arising from information contained within cross-correlations and higher-order statistics, it will be necessary to simultaneously model non-Newtonian, horizon-scale effects alongside subtle yet still important small-scale nonlinear physics.
Addressing this has been a focus of a number of studies, which have found different statistics and inferred cosmological parameters to be sensitive to such effects to various degrees \cite{1907.13109,2107.00467,Petri:2016qya,0910.3786,Munshi:2019csw,1101.4769}.

Accurately predicting observable properties of our Universe requires accurate knowledge of the spacetime metric through which information has propagated to us--in particular, knowledge of this metric on our past light cone. The idea of determining the metric on the past light cone from observations has been explored in a large-scale setting in several works \cite{Bester:2013fya, Bester:2016fbs, 1210.3069}.
In this work we introduce a novel method \textsc{LC-Metric} (Light Cone Metric restoration) for extracting the metric on our past light cone from either a mock density field on the lightcone, or a series of constant-time ``snapshots'' from simulations, and subsequently computing observable quantities.
We focus on the ability of this method to act as a post-processing procedure for the output of both standard Newtonian N-body simulations and cosmic emulators, especially those for which a small number of timesteps may have been taken and so recovery of the metric and its time derivative is less straightforward.
Once recovered, we can use the metric to subsequently account for gravitational effects in generated mock observations, including post-Born corrections to gravitational lensing and nonlinear contributions to the integrated Sachs-Wolfe (ISW) effect.

Nonlinear corrections to lensing and the ISW effects have been examined in literature using a variety of schemes in which approximations are employed to circumvent the need to determine the metric and its derivatives.
However, as accurately modeling observables will be important both at low and high redshift, on small and large scales, and for a number of observables including the cosmic microwave background (CMB) temperature and polarization, and galaxy power spectra, it will be important to consistently and correctly incorporate both nonlinear and large-scale lightcone-projection corrections \cite{1105.5292,1105.5280,1812.03167}.
The nonlinear ISW effect itself can be decomposed into several contributions, including the Rees-Sciama and moving-lens effects \cite{1003.0974}, which are expected to be detectable by future experiments with high signal-to-noise \cite{2006.03060}.
Gravitational lensing itself is perhaps better-studied in literature, including nonlinear contributions \cite{1101.4143, astro-ph/9901191,astro-ph/0101333,0809.5035,1202.2332,1312.1536,0711.1540,1702.03317,1904.07905,Lepori:2020ifz,Giblin:2017ezj}. Our approach in modeling these effects will be to first validate our approach, and then to produce lensing and ISW maps that are both consistent with other observables and also include nonlinear contributions.

For weak-lensing simulations, previous studies have employed an approximate thin-lens scheme using multiple planes or spheres of mass \cite{astro-ph/9610096,astro-ph/9901191,0809.5035,1202.2332,1312.1536,1606.01903,0711.3793,1210.3069,1702.03317,1904.07905}.
(See also \cite{1910.10625} for a comprehensive review and code comparison of some existing weak-lensing simulations.)
While some of these studies rely on the Born approximation, assuming unperturbed photon geodesics, other studies beyond the Born approximation find good agreement depending on the statistic and scales in question (\cite{0809.5035,1702.03317}).
In contrast to lensing, modeling the ISW effect requires knowledge of the time derivative of the metric potential, which makes its calculation more computationally challenging.
This is especially true when computing the ISW signal in the highly nonlinear regime, and when post-processing simulation data. Various techniques have been studied in past literature to this end \cite{Cai:2010hx,Carbone:2016nzj,Adamek:2019vko,Watson:2013cxa,Naidoo:2021ylw,Hassani:2020buk}. These all require either output from a large number of snapshots or on-the-fly raytracing, and so are not directly applicable as a post-processing step, which is especially important for fast emulators.

Here, we work to establish a numerical framework in which we can robustly compute nonlinear ISW and post-Born lensing contributions from Newtonian light cone simulation or emulator output.
Rather than considering specific effects or terms in such a calculation, we will aim to extract the full nonlinear ISW and lensing contributions.
We attempt to remain agnostic as to the simulation output; our method can be applied to a variety of sources, ranging from n-body codes such as Gadget \cite{astro-ph/0505010} to emulators that provide light cone  output such as L-PICOLA \cite{1506.03737}.
We will primarily rely upon the latter of these codes in this work, especially as output from COLA methods will converge to that of a standard N-body simulation in the limit of a large number of timesteps.
We compare our approach to conventional approximate approaches based on simulation snapshots or light cones, and compare the requirements for numerical convergence between our method and other approaches.
We find that our approach to modeling the weak-lensing power spectrum agrees well with other approaches that utilize the Born approximation, although we do note some differences.
Importantly, we find that our scheme significantly outperforms conventional schemes when modeling the ISW effect in the highly nonlinear regime.

We expect our results and framework to be relevant for modeling of higher-order and relativistic effects in cosmological observables such as the moving-lens and Rees-Sciama (nonlinear ISW) effects \cite{1003.0974}, higher-order contributions to observations of the CMB  including e.g. post-Born corrections or lensing of the SZ signal itself. While some of these corrections have been studied and may be detectable by upcoming experiments \cite{10.1093/mnras/stw2615, 1605.05662,2006.03060}, various higher-order effects have yet to be examined in detail--one of our goals is to initiate a study in this direction.
Our code has been made public and can be found at \footnote{https://github.com/ctian282/lcmetric}.

This paper is structured as follows. In Section \ref{sec:meth}, we review linearized GR theory, and describe our scheme to recast the linearized GR equations into a hyperbolic form on the past light cone.
We then introduce our numerical method for solving these equations, which employs a multigrid technique to relax these equations and determine the Newtonian potential and its derivatives.
In Section \ref{sec:res}, we calculate cosmological observables, including the weak-lensing and the (nonlinear) ISW effects, from the metric on the past light cone, and compare them with some conventional approximate methods. We summarize and conclude in Section \ref{sec:summ}.

\section{Methodology}
\label{sec:meth}

\subsection{Linearized GR equations}

The scheme we present here is based on a linearized treatment of the spacetime metric in Newtonian gauge (see e.g. \cite{Weinberg:2008zzc}). In this work, we will assume there is no anisotropic stress so the two Newtonian potentials are equal, although this assumption may be relaxed in a more general setting.
We further do not study the impact of vector or tensor perturbations, although these may be treated by a similar procedure.
Given these approximations, the perturbed metric in Newtonian gauge can be written as
\begin{equation}
\label{eq:13}
  \mathrm{d}s^{2} = a^2  \left[(1 + 2 \Phi) \mathrm{d}\tau^2 - (1 - 2 \Phi) \delta_{ij}(\mathrm{d}x^i\mathrm{d}x^j)\right],
\end{equation}
with $\Phi$ the Newtonian potential. The comoving Hubble parameter $\mathcal{H}$ is proportional to the Hubble parameter $H$ as $\mathcal{H} = a H$, and, in a Universe containing just dust and vacuum energy, it evolves according to the Friedmann equation
\begin{equation}
\label{eq:7}
\mathcal{H}^{2} = \mathcal{H}_0^2 a^{2}\left( \Omega_ma^{-3} + \Omega_{\Lambda} \right),
\end{equation}
where $\mathcal{H}_0$ represents the current expansion rate.
The background homogeneous stress tensor can be written
\begin{equation}
  \bar{T}_{\;\; \nu}^{\mu} = (\bar{\rho} + \bar{P})  \bar{U}^{\mu} 
  \bar{U}_v - \bar{P} \delta_{\nu}^{\mu},
\end{equation}
where $\bar{U}_{\mu} = a \delta^0_{\mu}$ and $\bar{U}^{\mu} = a^{-1} \delta_0^{\mu}$. The perturbation to the matter content $\delta T^{\mu}_{\nu}$ is 
\begin{align}
    \delta T^0_{\;\;0} &= \bar{\rho} \delta \\
    \delta T^i_{\;\;0} &= (\bar{\rho} + \bar{P}) v^i \\
    \delta T^0_{\;\;j} &= -(\bar{\rho} + \bar{P}) v_j \\
    \delta T^i_{\;\;j} &= -\delta P \delta^i_j + \Pi^i_{\;\;j},
\end{align}
where $\delta$ is defined $\delta \equiv (\rho - \bar{\rho}) / \bar{\rho}$ and $v_i$ donates the peculiar velocity field. 
Ignoring the anisotropic stress tensor $\Pi^i_{\;\;j}$ and keeping the linear terms in the metric and the matter field, the linearized Einstein equations can be written as
\begin{align}
  \nabla^2 \Phi - 3 \mathcal{H}  (\Phi' +
  \mathcal{H} \Phi)  &=  4 \pi Ga^2  \bar{\rho} \delta   \label{eq:9}\\
  \partial_i  (\Phi' + \mathcal{H} \Phi)  &=  - 4 \pi Ga^2  (\bar{\rho} +
  \bar{P}) v_i  \label{eq:10}\\
  \Phi'' + 3 \mathcal{H} \Phi' + (2 \mathcal{H}' + \mathcal{H}^2) \Phi  &=  4
  \pi Ga^2 \delta P  \label{eq:11}.
\end{align}
The prime symbol denotes a time derivative with respect to the conformal time $\tau$.
We set the pressure perturbation, $\delta P$, to zero under the collisionless-particle approximation in the remainder of the paper. 

\subsection{Connecting Newtonian N-body simulations to GR}

The vast majority of simulations of large-scale structure formations are performed within a Newtonian gravity framework, rather than a general relativistic setting.
While there are exceptions \cite{Adamek:2016zes,Giblin:2018ndw,Barrera-Hinojosa:2019mzo,Bentivegna:2015flc,Macpherson:2018btl,East:2019chx,Adamek:2017mzb}, in general this necessitates a translation between Newtonian simulation output and the corresponding general-relativistic quantities in the appropriate gauge.  One promising approach to reconciling this discrepancy involves re-interpreting the input and output to standard Newtonian codes in gauges specially developed for this purpose \cite{Giblin:2018ndw}.
Another straightforward option, which we employ here, is to transform the output of Newtonian codes to approximately agree with a relativistic interpretation as proposed in \cite{Chisari:2011iq}. The following dictionary maps output from Newtonian simulations to output in Newtonian gauge,
\begin{align}
\label{eq:15}
  \Phi &= \Phi_{\rm sim} \\
  \vec{v} &= \vec{v}_{\rm sim} \\
  \vec{x} &= \vec{x}_{\rm sim} + \delta \vec{x}_{\rm in}.
\end{align}
To find the value of corrections $\delta \vec{x}_{\rm in}$ to particle positions, we can simply solve the equation \cite{Chisari:2011iq}
\begin{equation}
\label{eq:17}
 \nabla \cdot \delta \vec{x}_{\rm in} = 5 \Phi_{\rm in},
\end{equation}
where $\Phi_{\rm in}$ is the gravitational potential determined when setting initial conditions. Our post-processing scheme uses this dictionary to correct particle positions given by N-body simulations. 
We also note that the (Newtonian) density contrast $\delta $ is subject to a GR correction 
\begin{equation}
\label{eq:16}
\delta = (1+3\Phi) \delta_{\rm sim} = (1+3\Phi)\bar{n} \sum_{i} \delta^{(3)}_{D}(\mathbf{r} - \mathbf{r}_{i}) - 1,
\end{equation}
where $\bar{n}$ is the average particle density,
and $\delta^{(3)}_D(\mathbf{r})$ is the $3$-d Dirac delta function.
In the test cases we present in this paper, the GR corrections are only appreciable at the largest scales.

\subsection{Solving Einstein's equations on the light cone}

Our past light cone can be thought of as a sequence of nested spherical shells. We will therefore work in a spherical polar coordinate system $(r, \theta, \phi)$, where $r$ is the comoving distance and $\theta$ and $\phi$ are the azimuthal and polar angles respectively. We further employ the coordinate transformations. 
\begin{eqnarray*}
  \tau & \rightarrow & \eta\\
  r + \tau & \rightarrow & w,
\end{eqnarray*}
under which varying $\eta$ corresponds to considering future/past null coordinate cones, while $w=\rm const$ corresponds to the surface of a specific null coordinate cone. We also manually set $w = 0$ for past light cone of the observer at $z = 0$.
The coordinate system after this transformation is equivalent to ``geodesic light cone  coordinates'' to leading order \cite{ELLIS1985315,Gasperini:2011us}.
To transform the equations of motion \Cref{eq:9,eq:10,eq:11} into the coordinate system $(\eta, w, \theta, \phi)$, we can rewrite the EOM  \Cref{eq:9,eq:10,eq:11} in standard spherical coordinates $(\tau, r, \theta, \phi)$ and use the following transformation relations for partial derivatives 
\begin{eqnarray*}
  \partial_{\tau} & \rightarrow & \partial_{\eta} + \partial_w\\
  \partial_r & \rightarrow & \partial_w\\
  \partial_{\tau}^2 & \rightarrow & \partial_{\eta}^2 + 2 \partial_{\eta}
  \partial_w + \partial_w^2\,\\
  \partial_r^2 & \rightarrow & \partial_w^2.
\end{eqnarray*}
These are derived directly from the coordinate transformations relations, and can be used to rewrite the EOMs in the new coordinate system,
\begin{align}
  \ddot{\Phi} & = - \frac{2}{r} \Pi + 2 \mathcal{H} \Pi - 2 (\dot{\mathcal{H}}
                - \mathcal{H}^2) \Phi + 4 \pi Ga^2  \bar{\rho} \delta  \nonumber \\
  &+ 3 \times (4 \pi Ga^2
  \bar{\rho} \Phi) - \frac{\nabla^{(2)} \Phi}{r^2} + 2 \times 4 \pi Ga^2 
  \bar{\rho} v_r  \label{eq:28}\\
  \dot{\Xi} & = -2 \mathcal{H} \Xi - \mathcal{H} \Omega - (2\dot{\mathcal{H}} + \mathcal{H}^{2}) \Phi + 4 \pi Ga^2  \bar{\rho} v_r .
  \label{eq:30}
\end{align}
Here, we have defined some shorthands for partial derivatives in the new coordinates,
\begin{align}
  \label{eq:dir_short}
\dot{X}\equiv\frac{\partial X}{\partial \eta},\;\; X'\equiv\frac{\partial X}{\partial r} = \frac{\partial X}{\partial w}
\end{align}
for a field $X$,
and also shorthands for several new variables,
\begin{align}
  \label{eq:var_short}
\Omega \equiv  \dot{\Phi},\;\; \Pi \equiv \Phi',\;\; \Xi \equiv \partial_{\tau}\Phi =  \Pi + \Omega.
\end{align}
Eqs.~\eqref{eq:28} and \eqref{eq:30} comprise a second-order linear PDE coupled to a first order PDE. We discretize the solution onto a spherical mesh by discretizing the comoving time direction $\eta(\tau)$ uniformly, and discretizing angular directions $(\theta,\phi)$ according to the {HEALPix} convention. Under this discretization, the density contrast $\delta$ and the radial velocity fields $v_{r}$ can be extracted from particle data using any preferred deposition scheme, such as Nearest-Grid-Point (NGP) or Cloud-in-Cell (CIC) deposition. The weights we use for radial grids correspond to standard CIC weights, while the angular weights are given by the {HEALPix} \texttt{get\_interpol} function. To calculate the angular Laplacian term $\nabla^{(2)} \Phi$ in Eq.~\eqref{eq:28}, 
we also employ {HEALPix} to perform spherical harmonics transformations to compute the spherical harmonic coefficients $a_{lm}$ for all terms in the Eqs.~\eqref{eq:28} and \eqref{eq:30} under the convention
\begin{align}
  \label{eq:spht}
a_{lm} = \frac{4\pi}{N_{\rm pix}} \sum_{p=0}^{N_{\rm pix}-1}Y_{lm}^{*}(\bm{\theta}_p)f(\bm{\theta}_{p})\,.
\end{align}
In harmonic space,  $\nabla^{(2)} \Phi$ can be simply represented by $\ell(\ell+1) \Phi_{lm}$, where $\Phi_{lm}$ are the spherical harmonic coefficients of $\Phi$.
For radial derivatives in \Cref{eq:28,eq:30}, we employ a 2nd-order finite-differencing method to calculate them. 

Two boundary conditions for $\Phi$ (cf. Eq.~\eqref{eq:28}) and one boundary condition for $\Xi$ (cf. Eq.~\eqref{eq:30}) are also required. The boundary conditions for $\Phi$ can be extracted from initial (higher $z$) and final (lower $z$) snapshot data,  interpolating the $\Phi$ field on these slices to recover corresponding values on the {HEALPix} grids. The initial condition for $\Xi$ can be estimated from the same initial snapshot with the knowledge of the velocity field \cite{Cai:2010hx}.
Specifically, combining the Friedmann equation Eq.~\eqref{eq:7} with the Poisson equation for gravity, in Fourier space, the Newtonian gravitational potential can be written as
\begin{equation}
\label{eq:19}
\Phi(\bm{k}, \tau) = -\frac{3}{2}\left(\frac{H_{0}}{k}\right)^{2} \Omega_m \frac{\delta(\bm{k}, \tau)}{a}.
\end{equation}
By plugging-in the continuity equation $\dot{\delta}(\bm{k},\tau) + i \bm{k} \cdot \bm{p}(\bm{k}, \tau)=0$ in Fourier space, we have
\begin{equation}
\label{eq:20}
\frac{\partial\Phi}{\partial\tau}(\bm{k},\tau) = -\frac{3}{2}\left(\frac{H_{0}}{k}\right)^{2} \Omega_m \left[\frac{H}{a} \delta(\bm{k}, \tau) +  \frac{i \bm{k} \cdot \bm{p}(\bm{k}, \tau)}{a}\right],
\end{equation}
where $\bm{p}(\bm k, \tau)$ is the Fourier transform of $\left[1+\delta(\bm{x}, \tau)\right] \bm{v}(\bm{x}, \tau)$, and the velocity field $\bm{v}(\bm{x}, \tau)$ is again extracted from the CIC or NGP scheme on the initial snapshot.

\begin{figure}[ht]
  \centering
  \includegraphics[width=0.44\textwidth]{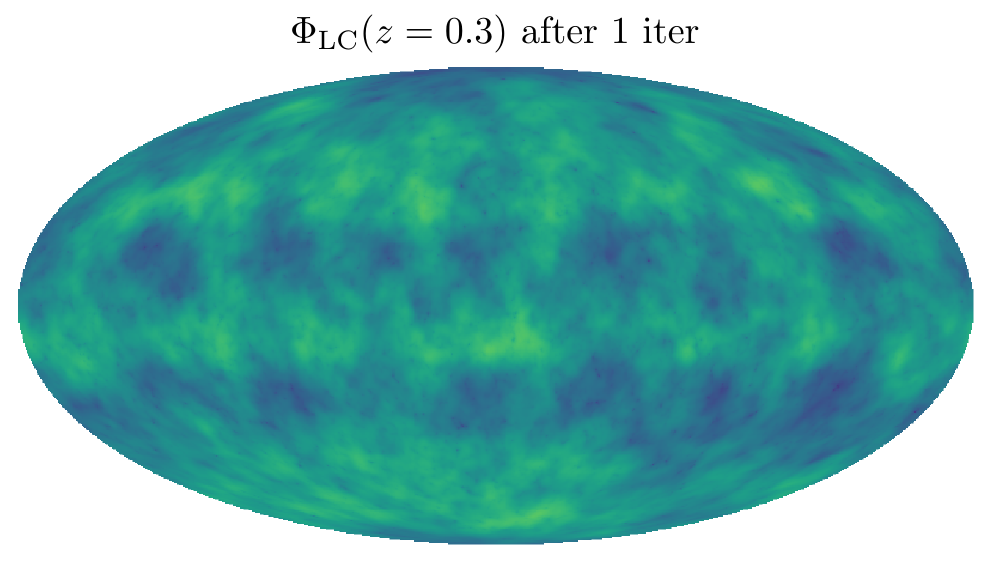}
  \includegraphics[width=0.44\textwidth]{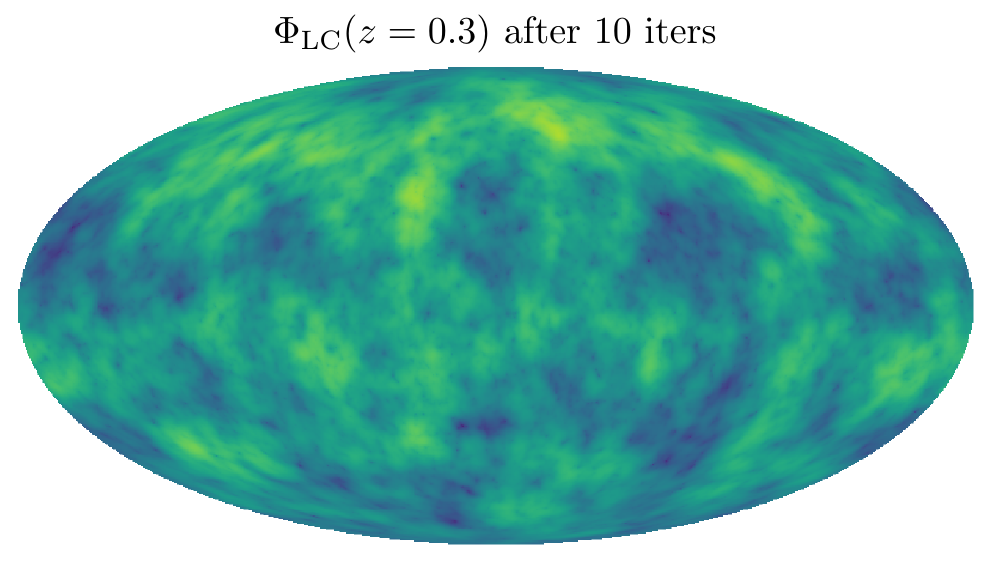}
  \includegraphics[width=0.44\textwidth]{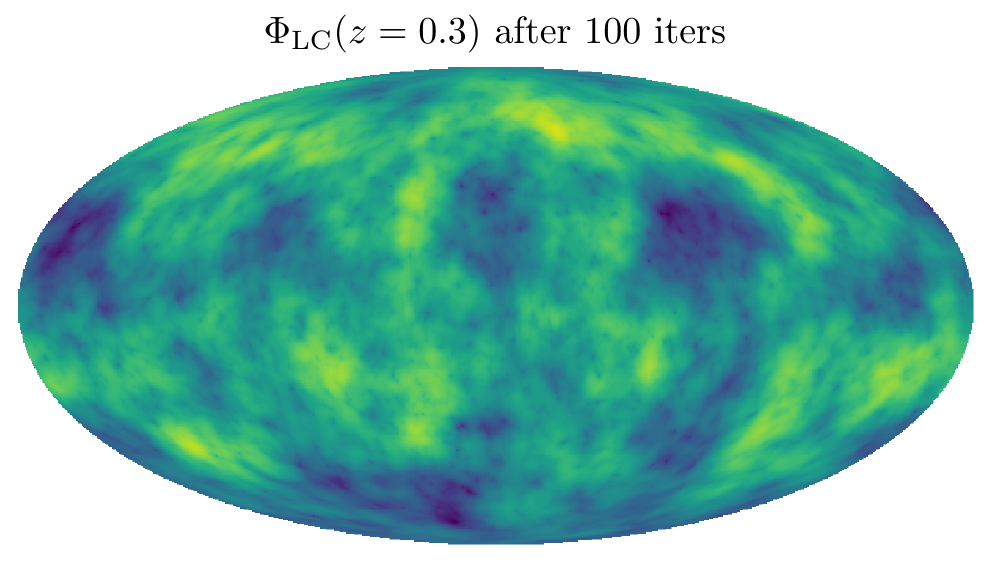}
  \includegraphics[width=0.44\textwidth]{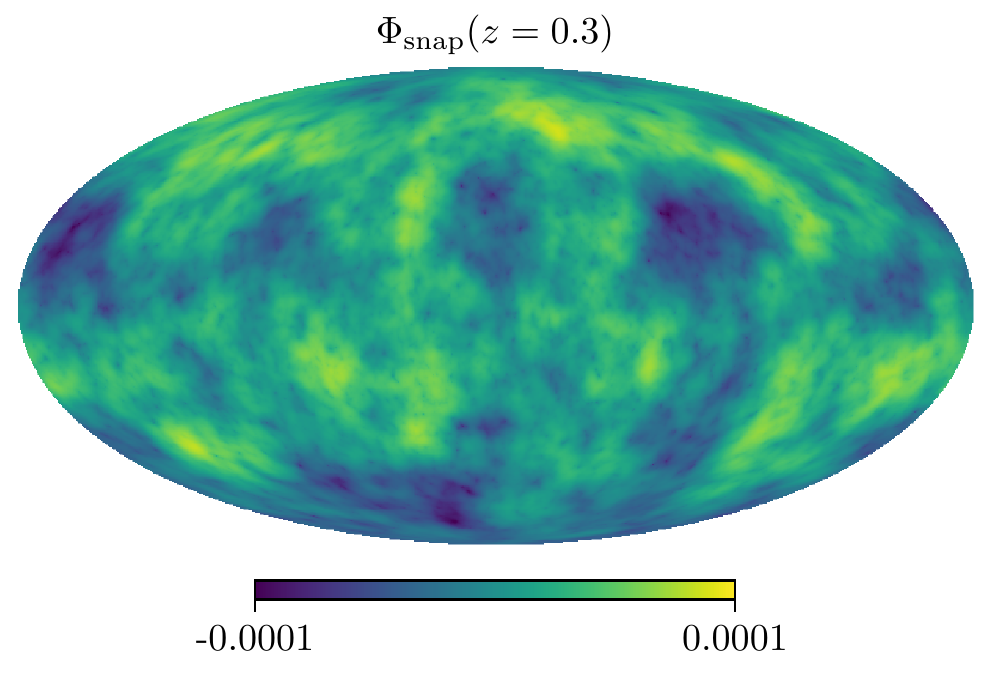}
  \caption{\label{fig:cmp_Phi} Comparison of a $\Phi$ snapshot at $z=0.3$ between the direct snapshot output $\Phi_{\rm snap}$ and the restored $\Phi_{\rm LC}$ after 1, 10, and 100 iterations (No. of   V-cycles).}
\end{figure}

After imposing boundary conditions at higher and lower redshifts \footnote{When solving Eq.~\eqref{eq:30}, we only need to impose a boundary condition at high redshift}, we use a relaxation scheme to solve for each spherical harmonic coefficient in each radial bin.
When the size of voxels on spherical grids approximately matches the voxels from Cartesian grids of N-body simulations, using too many spherical grids will prevent an ordinary relaxation technique (e.g. Newton's method) from converging to the true solutions of \Cref{eq:28,eq:30} in a practical amount of time.
We therefore employ a multi-grid method \cite{FEDORENKO19621092}, which builds a hierarchy of the computational grids to accelerate the relaxation procedure. We find that employing the multi-grid method will result in approximately a 10x speedup compared to the conventional Newton relaxation method. The relaxation process is accomplished by repeating V-cycle iterations \cite{10.5555/1403886}, and an example for the relaxation procedure for the reconstructed potential $\Phi_{\rm LC}$ at $z = 0.3$ is shown in Fig.~\ref{fig:cmp_Phi}. The $\Phi_{\rm LC}$ quickly converges to the potential on the snapshots $\Phi_{\rm snap} $ at the same redshift as more and  more  V-cycle  iterations  are  performed. Further details about the convergence of the solver can be found in Appendix \ref{subsec:solver}.

\subsection{Simulation Data and light cone  construction}
\label{subsec:data}

\begin{figure*}[ht]
  \centering
  \includegraphics[width=0.82\textwidth]{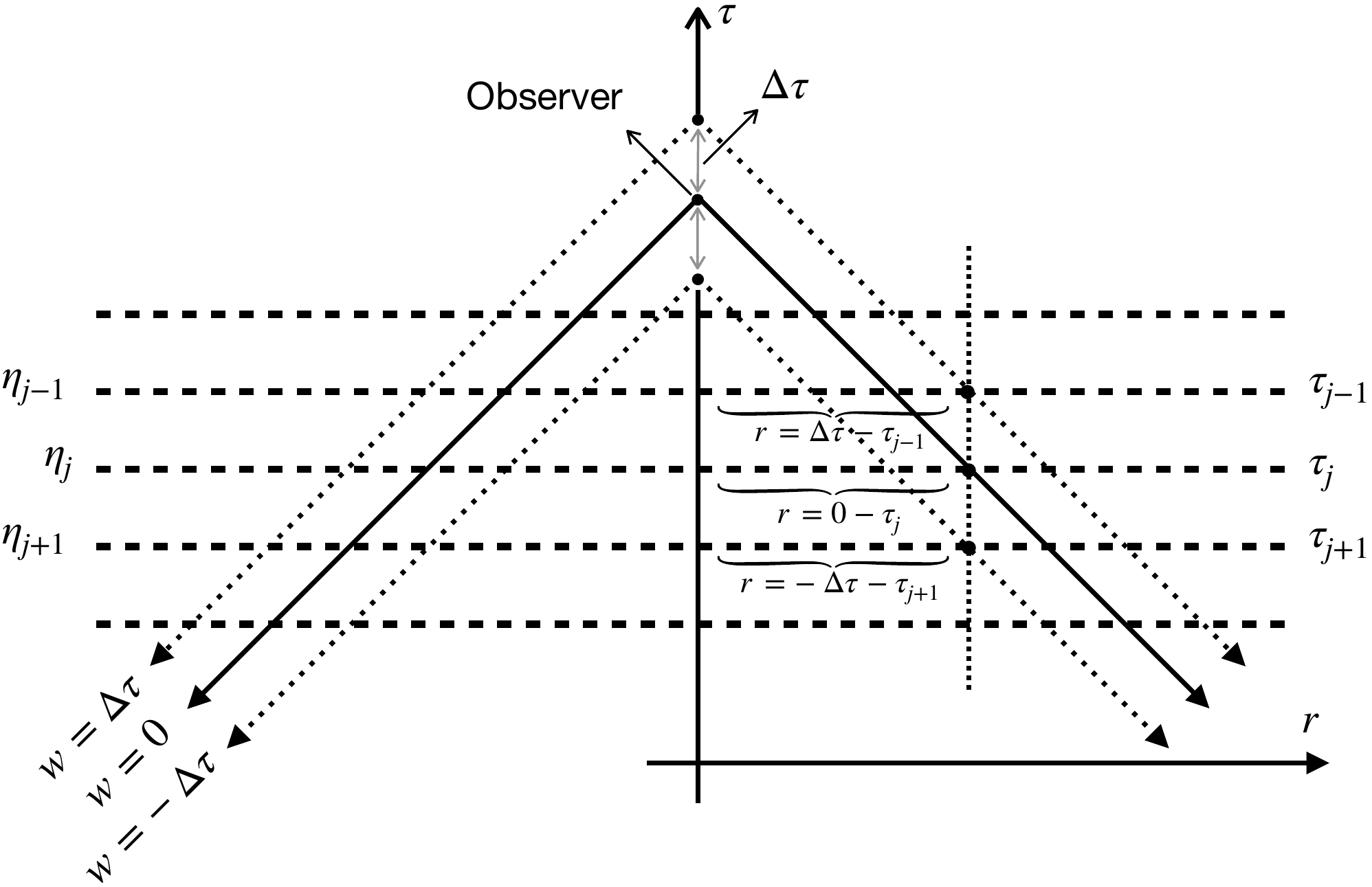}
  \caption{\label{fig:LCs} Illustration of the construction of ``thick'' light cones, which are the light cones containing information between the dotted lines in the figure. They are computed by obtaining additional light cone  data for observers separated positively and negatively in time by a conformal time $\Delta\tau$, which is also equal to the radial resolution of the light cone  mesh. The 3 dots representing 3 mesh-grids on each light cone as an example thus have equal spatial coordinates.}
\end{figure*}

To validate our metric reconstruction procedure, we will focus on post-processing simulated light cone data, which is most accurately constructed through on-the-fly interpolation during N-body simulations.
However, as the most common output format of N-body simulations is snapshots, our scheme is also designed to be able to work with light cone data restored from snapshots.
To validate the \textsc{LC-Metric} reconstruction scheme, we run a suite of dark-matter-only N-body simulations using the L-PICOLA code \cite{Howlett:2015hfa}, and subsequently use \textsc{LC-Metric} to post-process its output.
L-PICOLA implements the COLA method, a reference formulation using 2nd-order Lagrangian perturbation theory to quickly solve the Newton-Vlasov system \cite{Tassev:2013pn}.
A significant speed-up compared to conventional particle-mesh method is typically found, especially on large scales; further, this method will produce equivalent results to standard N-body codes for sufficiently small timesteps \cite{Howlett:2015hfa}.

Our simulations are carried out with $1024^3$ dark-matter particles in a periodic box with side length $L=1024h^{-1}\,\rm Mpc$,
corresponding to a particle mass of approximately $10^{11} M_{\odot}$.
The density field of dark-matter particles is CIC-deposited onto $1024^3$ Cartesian grids, and forces are computed using Fast Fourier Transforms.
The background cosmology is chosen to be a flat $\Lambda \rm CDM$ model with parameters $\Omega_m=0.31$, $\Omega_{\Lambda}=0.69$, $ h=0.69$, $\sigma_8=0.83$, and $n_{s}=0.96$.
Initial conditions for the simulations are generated based on a linear transfer function computed using CLASS with matching cosmological parameter values \cite{Blas:2011rf} at $z=9$.

As the \textsc{LC-Metric} scheme supports both light cone N-body data and snapshots data, we run the L-PICOLA code twice under light cone mode and normal (snapshot) mode to generate corresponding outputs with the same random seed. When snapshots data is given, we construct the light cone data from the snapshots by extrapolating particle positions according to their velocities and calculating their intersections with the light cone. We also replicate our computational box to cover the entire light cone.
Although this replication procedure will result in artifacts on large scales, as a proof-of-concept for our metric-recovery scheme, we will primarily focus on whether the mock light cone metric can be correctly computed from a given density field. 
Further details of our scheme for constructing light cone data from snapshots is described in Appendix \ref{sec:lc_from_snaps}.
By choosing the two boundaries slices to be at redshifts $z\approx 0.03$ and $z \approx 0.52$, we construct our light cone metric extending between these redshifts.  
Note that even though we test our code by post-processing L-PICOLA N-body output, our scheme will work with generic light cone or snapshot data  that contains particle positions and peculiar velocities.

When constructing a light cone from snapshots, a useful technique supported by \textsc{LC-Metric} is the ``thick" light cones scheme. As shown in Fig.~\ref{fig:LCs}, by shifting the original light cone in the time direction, we construct two auxiliary light cones. The amount by which the light cone is shifted is set to be equal to the radial resolution of light cones, so that the voxels both have the same conformal radius $r$, and are aligned in the time direction. Under this configuration, the partial derivative with respect to conformal time of any field can be calculated directly using the 2nd-order finite-differencing scheme. The ``thick" light cones scheme will also facilitate the calculation when light-rays deviate from the ``thin" light cone when tracing the null geodesics precisely.

After reading in light cone data, we deposit particle masses and radial peculiar velocities onto our spherical grids, which have radial resolution $N_{\rm \eta} = 1024$ and angular resolution corresponding to $N_{\rm side} = 512$.
The number of radial grids is chosen to ensure the radial resolution of the light cone $ {\rm \Delta} \eta = {\rm \Delta} r$ approximately matches the Cartesian resolution ${\rm \Delta}x$; and the angular resolution is set to make sure that most of the spherical-coordinate voxels are larger than the Cartesian voxels to avoid oversampling. Appendix \ref{sec:conv} discusses how the cosmological observables vary with the choice of resolution.

We will further need to account for power suppression due to the implicit convolution 
associated with depositing particle masses onto grids (either through CIC or NGP). We deconvolve the grids using corresponding window functions to restore a true representation of the density field. For N-body simulation data from L-PICOLA, we use a window function \cite{1981csup.book}
\begin{equation}
\label{eq:21}
W(\bm k) = \left[\mathrm{sinc}\left(\frac{\pi k_{x}}{2 k_{N}}\right) \mathrm{sinc}\left(\frac{\pi k_{y}}{2 k_{N}}\right)
  \mathrm{sinc}\left(\frac{\pi k_{z}}{2 k_{N}}\right) \right],
\end{equation}
where $k_N$ is the Nyquist frequency, and the underlying density field $\delta^{m}(\bm k)$ is calculated from
\begin{equation}
\label{eq:22}
\delta^{m}(\bm k) = W(\bm k)^{-p} \delta(\bm k ),
\end{equation}
where $\delta(\bm k)$ is the density field obtained from mass depositing scheme directly, and p is $1$ and $2$ for NGP and CIC scheme respectively.

Similar to the NGP or the CIC particle-deposition procedure on Cartesian grids, when depositing particle masses onto the {HEALPix} grids, a similar window function $W_{\ell}\approx \mathrm{sinc}(\ell \Delta \theta/2\pi)$ should be considered, and the density fields on the HEALPix grids also need to be corrected by
\begin{equation}
\label{eq:23}
\delta^{m}_{l} = W_{l}^{-p} \delta_{l},
\end{equation}
where $p$ is similarly $1$ and $2$ for the NGP and CIC depositing scheme. In this study, we choose to scale all the spherical-harmonics coefficients in the density and velocity fields according to the approximate window function available through {HEALPix}.

A smoothing procedure is also introduced for the density fields on HEALPix grids through the size of individual HEALPix pixels possibly being significantly smaller than the resolution of an N-body simulation at low comoving radius. Unless specified, we convolve the density and velocity fields on HEALPix grids with a Gaussian smoothing kernel with half-max width
\begin{equation}
\label{eq:24}
\theta_s = \Delta x / 2r ,
\end{equation}
where $\Delta x $ is the resolution of the N-body simulation and $r$ is the comoving radius at each specific spherical shell. The inner shell (at lower redshift) will have a larger smoothing angle due to the limited Cartesian resolution of the N-body simulations.

\subsection{Theory of Lensing and ISW effects}
\label{subsec:lensing_and_ISW}

To verify our metric-reconstruction strategy, we generate and examine cosmic observables, including the lensing-convergence and (nonlinear) ISW effects.

Weak gravitational lensing by large-scale structure is due to the gravitational deflection of light by intervening matter as it travels to us. In this paper, we will focus on quantifying the weak-lensing convergence
\begin{equation}
\label{eq:2}
\kappa \equiv \frac{D_A - \bar{D}_{A}}{\bar{D}_A},
\end{equation}
where $\bar{D}_A$ is the unperturbed angular-diameter distance in an FLRW background. Under the assumption of unperturbed photon trajectories, the so-called Born approximation, the lensing potential $\psi(r_s, \bm{\theta})$ depending on the gravitational potential $\phi$ for sources at comoving distance $r_s$ (from us observers) can be defined as
\begin{equation}
\label{eq:1}
\psi(r_s, \bm{\theta}) \equiv - \int_0^{r_s} d r' \frac{r_{s} - r'}{r_{s} r'}\times 2\Phi(r', \bm{\theta}),
\end{equation}
and the lensing convergence is then
\begin{equation}
\label{eq:3}
\kappa = -\frac{1}{2} \nabla^{(2)} \psi.
\end{equation}

To avoid the need of exactly estimating the lensing potential in Eq.~(\ref{eq:1}), many studies choose to introduce the radial modes and replace the angular Laplacian in Eq.~(\ref{eq:2}) with the 3-d Laplacian $\nabla^2 \Phi$. Then following the Newtonian Poisson equation
\begin{equation}
    \nabla^2 \Phi  =  4 \pi Ga^2  \bar{\rho} \delta,
\end{equation}
Eq.~(\ref{eq:2}) can be cast into an integration of the density field as
\begin{equation}
\label{eq:kappa_delta}
    \kappa =  \int_0^{r_s} d r' \frac{r_{s} - r'}{r_{s} r'} \times 4\pi G a^2 \bar{\rho}\delta(r', \bm{\theta}).
\end{equation}
In practice, a more widely used scheme--the ``thin-lens approximation''--relies on summing over light cone particles along lines of sight within each sky pixel.
This scheme is based on Eq.~(\ref{eq:kappa_delta}),  but does not require radial binning or extracting the gravitational potential, and has been employed in a number of contexts in the literature \cite{0807.3651,Kiessling:2010uc,Cayuso:2018lhv,Takahashi:2017hjr,Izard:2017kma},
\begin{equation}
\label{eq:4}
\kappa =\frac{2}{3} H_{0}^2 \Omega_{m,0} \frac{V_{\rm sim}/N_{\rm sim}}{\Omega_{\rm pix}} \sum_{\rm particles} \frac{1}{r_pa(r_p)} \frac{r_{s} - r_p}{r_s}.
\end{equation}
Here, $r_p$ is the comoving distance of the light cone particles and $\Omega_{\rm pix}$ is the size of the solid angle of sky pixels. Note that even though the approximation of introducing the radial modes holds accurately on small angular scale \cite{astro-ph/9901191}, its validity is not guaranteed on largest scales. 

Photons traveling through large-scale structure also encounter a time-varying potential--the ISW effect. For CMB photons, this effect alters their frequency, and hence the CMB temperature $T$, according to
\begin{equation}
\label{eq:6}
\frac{\Delta T_{\rm ISW}(r, \bm{\theta})}{T_{\rm CMB}} = 2 \int \frac{\partial \Phi(r, \bm{\theta}, \tau)}{\partial \tau} \mathrm{d}\tau\,.
\end{equation}
One could employ Eq.~(\ref{eq:20}) to estimate this ISW signal as suggested in \cite{Cai:2010hx}. If no knowledge of velocity fields in Eq.~(\ref{eq:20}) is given, one could assume the linear growth of $\delta$ then the $\partial \Phi(r, \bm{\theta}, \tau) / \partial \tau$ field. However, the linear approach fails to capture all nonlinear effects, which soon become significant on angular scales corresponding to the multipole mode around $l \sim 60$, and will dominate the ISW contribution beyond $l \sim 100$ \cite{Cai:2010hx,Adamek:2019vko}.

\section{Results}
\label{sec:res}

In this section, we demonstrate this new scheme's ability to reconstruct the spacetime metric as well as the accuracy with which we can compute lensing and ISW observables.
We provide a comparison of our scheme  both to the output of Einstein-Boltzmann solvers (CLASS) and to conventional methods found in past literature.
We explore the requirements for each of these different techniques and show that for the weak-lensing and (nonlinear) ISW observables, a better than than $5\%$ accuracy for $\ell \lesssim N_{\rm side}$ is achieved.

\subsection{Direct potential comparison}
\label{subsec:comp}

\begin{figure}[ht]
  \centering
  \includegraphics[width=0.48\textwidth]{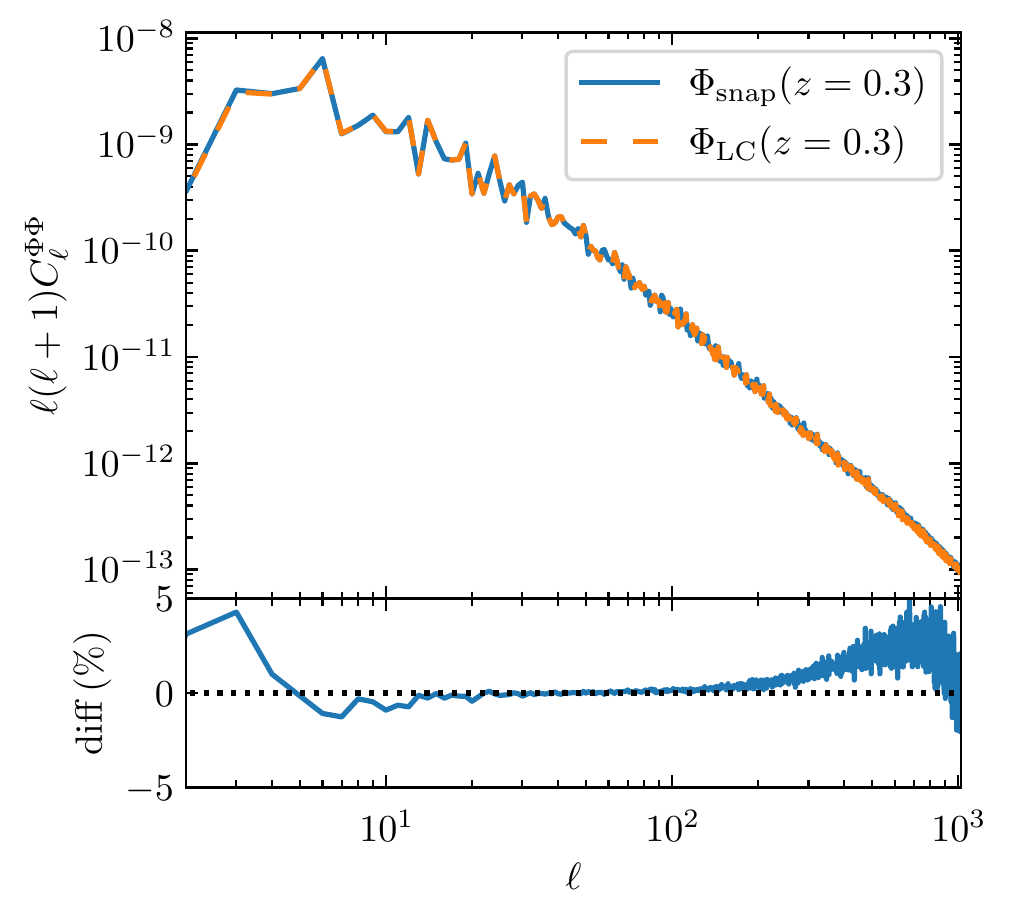}
  \caption{\label{fig:Phi_pow} Upper panel: Comparison of the angular power spectrum at of $\Phi$ at $z = 0.3$ between snapshot output $\Phi_{\rm snap}$ and the reconstructed potential $\Phi_{\rm LC}$ from the \textsc{LC-Metric} based on the light cone. Lower Panel: percent difference between $\Phi_{\rm snap}$ and $\Phi_{\rm LC}$. Convergence between the two methods is found as numerical resolution is increased.}
\end{figure}

\begin{figure}[ht]
  \centering
  \includegraphics[width=0.48\textwidth]{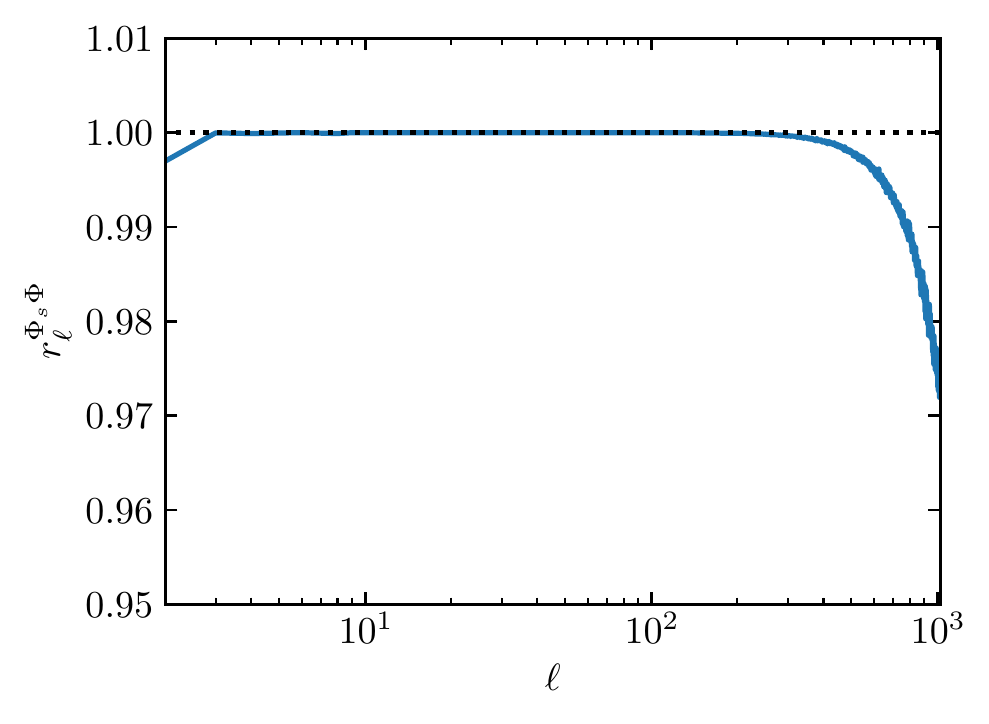}
  \caption{\label{fig:phi_cross} Normalized cross-correlation coefficients (Eq.~(\ref{eq:ncc})) between the restored potential $\Phi_{\rm LC}$ and the potential on the snapshot $\Phi_{\rm snap}$ at $z = 0.3$.  }
\end{figure}

We first evaluate the accuracy of our scheme by directly comparing the light cone-reconstructed gravitational potential $\Phi$ with the potential obtained from snapshot data. For the first such comparison,  we adopt the same configuration as in Fig.~\ref{fig:cmp_Phi}. The potential $\Phi_{\rm snap}$ is computed and interpolated to HEALPix grids from a snapshot at $z = 0.3$, while the reconstructed $\Phi_{\rm LC}$ is computed using the \textsc{LC-Metric} scheme under CIC deposition, using light cone data in addition to two spatial boundary slices at $z\approx 0.03$ and $z\approx 0.52$. No snapshot information at any redshift in between is used.
In Fig.~\ref{fig:Phi_pow}, we calculate the angular power spectra for $\Phi_{\rm snap}$ and $\Phi_{\rm LC}$. We also show their normalized cross-correlation coefficients defined as 
\begin{equation}
\label{eq:ncc}
    r_{\ell}^{XX'} = C_{\ell}^{XX'} / \sqrt{C_{\ell}^{X X} C_{\ell}^{X' X'}} 
\end{equation}
in Fig.~\ref{fig:phi_cross}, where $C_{\ell}^{XX'}$ is the cross-power spectrum for arbitrary fields $X$ and $X'$. 

According to Fig.~\ref{fig:Phi_pow} and Fig.~\ref{fig:phi_cross}, highly correlated signals are in percent-level agreement for $5 \lesssim \ell \lesssim N_{\rm side} $, with an accuracy better than $0.5\%$ for $20 \lesssim \ell \lesssim 200$.
The $5\%$ discrepancy for $\ell \lesssim 5$, much less than the sample variance for this range of multipoles, is likely due to the approximation used by the L-PCOLA code to generate the light cone data, and to noise arising from depositing particles masses and velocities to a spherical grid.
The accuracy at larger multipoles $\ell > N_{\rm side}$ is limited by the precision of the interpolation operations (see Appendix \ref{subsec:interp_conv} for further discussion).
The convergence test (see Appendix \ref{subsec:phi_conv}) also suggests that we can expect an improved reliability when the angular resolution $N_{\rm side}$ is increased.

\subsection{Weak Lensing}
\label{subsec:lensing}

The weak-lensing convergence $\kappa$ can be estimated in a number of different ways for sources at the high-redshift end of the light cone ($z \approx 0.48$). The thin-lens scheme (Eq.~\ref{eq:4}) provides one option that avoids the need to decompose the lensing (mass) distribution into radial bins at the cost of ignoring the radial modes. Another option is to compute the lensing convergence explicitly once the metric is known (\Cref{eq:3,eq:1}) on the light cone. 

\begin{figure}[ht]
  \centering
  \includegraphics[width=0.48\textwidth]{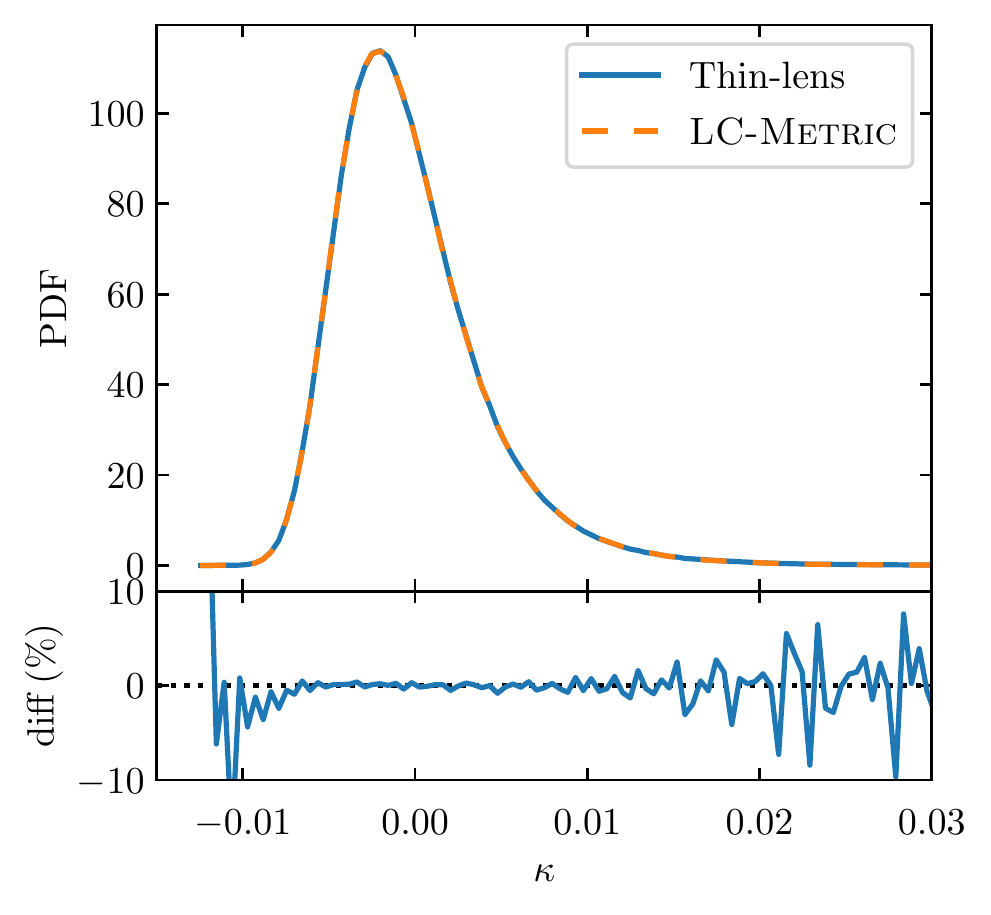}
  \caption{\label{fig:lensing_pdf}  Upper panel: Probability density functions (PDF) of the lensing convergence from $z=0.03-0.48$ calculated using the thin-lens approximation (Eq.~(\ref{eq:4})) and the lensing potential (Eq.~(\ref{eq:3})) based on the reconstructed metric. The mean value has been subtracted. Lower Panel: percent difference in the PDF between the thin-lens and the \textsc{LC-Metric} scheme.}
\end{figure}

We compare the probability density functions (PDF) of the lensing convergence computed using both the thin-lens approximation and explicit calculation from light cone-reconstructed metric potential, shown in Fig.~\ref{fig:lensing_pdf}. The PDFs agree to $0.5\%$ in the range of $\kappa$ where the PDFs are appreciably different than zero. The reconstructed $\Phi_{\rm LC}$ in this comparison is obtained from light cone data with NGP mass deposition scheme in order to match the thin-lens scheme Eq.~(\ref{eq:4}), which effectively employs a NGP method. Similar to the \textsc{LC-Metric} scheme, the corresponding window function $W_{l}$ (see Eq.~(\ref{eq:23})) is also applied to correct the power suppression after the depositing scheme is applied in the thin-lens procedure.

\begin{figure}[t]
  \centering
  \includegraphics[width=0.48\textwidth]{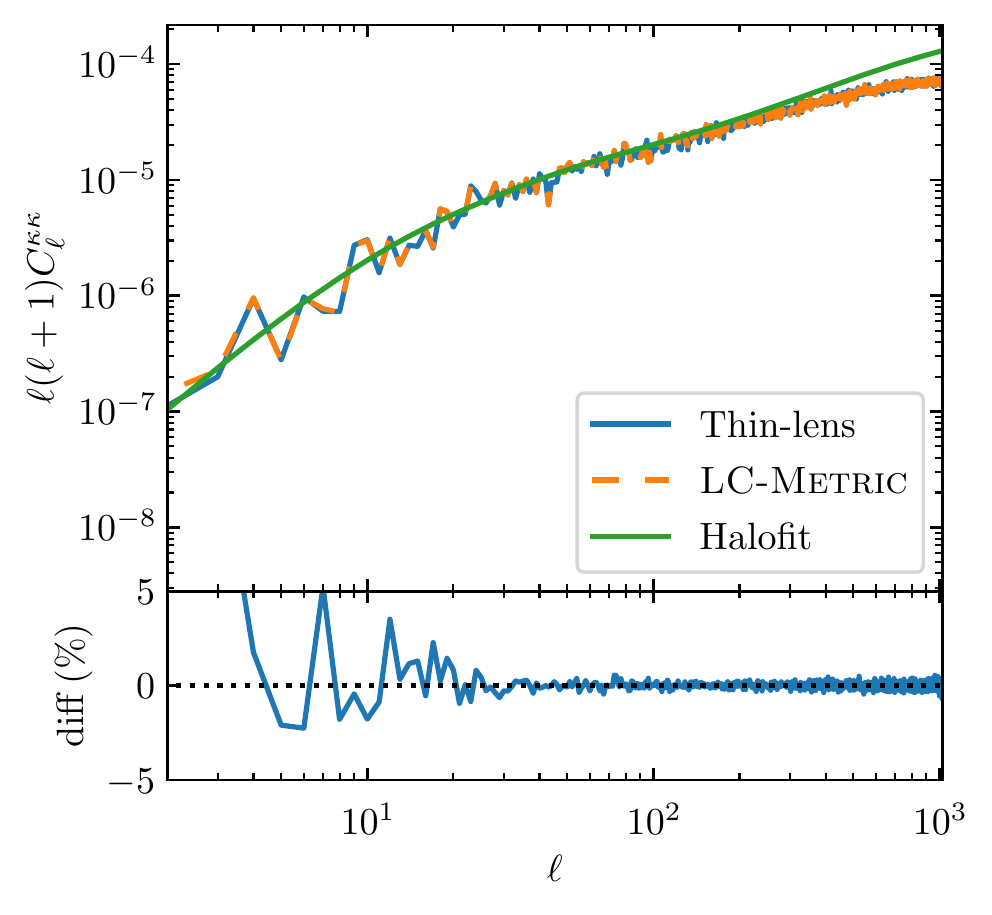}
  \caption{\label{fig:lensing_pow} Upper panel: Power spectrum of the lensing convergence $\kappa$ calculated with the thin-lens scheme (Eq.~(\ref{eq:4})) and the lensing potential (Eq.~(\ref{eq:3})) based on the reconstructed light cone metric $\Phi_{\rm LC}$ from the \textsc{LC-Metric}. Lower panel: percent difference between the thin-lens and the \textsc{LC-Metric} scheme.}
\end{figure}

In addition, the lensing power spectra calculated for each of these two methods are also shown in Fig.~\ref{fig:lensing_pow}, alongside a baseline power spectrum given by Halofit \cite{Smith:2002dz}. They are both consistent with the Halofit prediction over a wide range of scales. The discrepancy at higher $\ell$ is due to insufficient resolution in the N-body simulation and light cone grids, and is also observed in similar weak-lensing investigations \cite{Lepori:2020ifz,Liu:2017now}.   Only a sub-percent discrepancy is found for $\ell > 20$ between the thin lens and the \textsc{LC-Metric} based scheme, suggesting the \textsc{LC-Metric} reconstruction scheme is able to reproduce results that are fully consistent with the conventional thin lens approximation. The discrepancy at lower $\ell$ is revealing the errors of ignoring the radial modes in Eq.~(\ref{eq:kappa_delta}) and subsequently, Eq.~(\ref{eq:4}). These radial modes turn out to have a few percent effects for the convergence at very low multipoles. 

\begin{figure}[ht]
  \centering
  \includegraphics[width=0.48\textwidth]{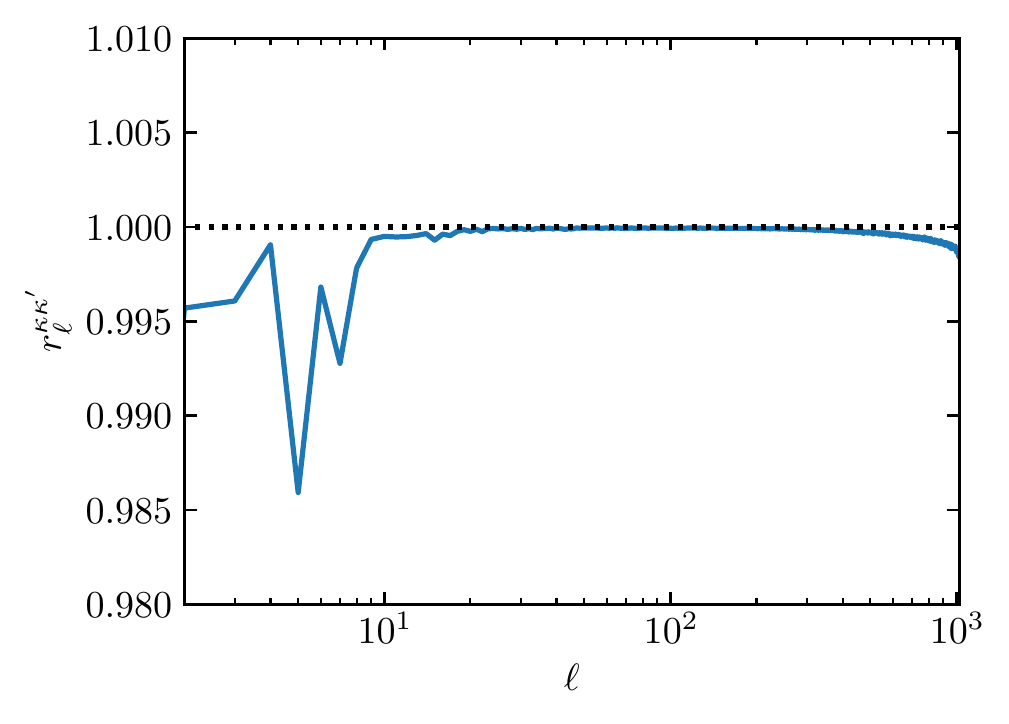}
  \caption{\label{fig:lensing_cross} Normalized cross correlation coefficients (Eq.~(\ref{eq:ncc})) of the lensing convergence $\kappa$ between the thin-lens scheme (Eq.~(\ref{eq:4})) and the lensing potential (Eq.~(\ref{eq:3})) based on the reconstructed light cone metric $\Phi_{\rm LC}$.}
\end{figure}

The normalized cross-correlations coefficients (Fig.~\ref{fig:lensing_cross}) confirm the strong agreement between the two methods. The \textsc{LC-metric} scheme, however, can take advantage of the full metric information on the past light cone, allowing for the study of higher order effects beyond the Born approximation through a post-Born ray tracing. We leave the exploration of such effects to a future study.

\subsection{Nonlinear ISW}
\label{subsec:isw}

The ISW effect involves the integration of the time derivative of $\Phi$ ($\Xi$ field), which can be decomposed into $\partial_{\eta} + \partial_{w}$ in our coordinate system, see Eq.~(\ref{eq:30}).
However, because the power of the nonlinear ISW effect (including moving lens and Rees-Sciama effects) decays quickly at higher $\ell$, high-frequency noise due to the imperfect estimation of the velocity field in this regime can easily surpass the signal. We therefore employ ``thick'' light cones introduced in Section \ref{subsec:data}, relying on the light cone data constructed from snapshots to calculate the $\partial \Phi/\partial \tau$ term using a finite differencing scheme. 
We truncate the $\partial \Phi / \partial \tau$ field near the boundaries (at initial and final redshift) of the light cone to avoid boundary effects. The redshift range of the ISW effect we model in this work, following the truncation, ranges from $z \approx 0.07 - 0.47$. The snapshots used in our constructions are uniformly distributed in the space of co-moving distance.

To accurately compute the ISW contributions in the highly nonlinear regime, we also implement and examine an ``on-the-fly'' prescription. 
We re-run our N-body simulations with the same parameters and a much finer time step (1024 steps).
We again calculate the $\partial \Phi / \partial \tau$ using a finite differencing scheme by subtracting snapshot data on current step from the previous one. We then interpolate the $\partial\Phi / \partial \tau$ data from the Cartesian grids to our HEALPix grids, and finally, we integrate all the $\partial\Phi / \partial \tau$ data at each HEALPix pixel to evaluate the term Eq.~(\ref{eq:6}).
Because of the number of time steps required to ensure integration convergence (see Appendix \ref{subsec:isw_conv}), this scheme can be taken as an accurate reference of the nonlinear ISW, though it is not an especially efficient technique for processing simulation data due to the resolution and time-stepping requirements. 

We have also implemented the method described in \cite{Cai:2010hx}, which estimates $\partial\Phi / \partial \tau$ using Eq.~(\ref{eq:20}) on every output snapshot. After obtaining all the $\partial\Phi / \partial \tau$ data, similar to \cite{Cai:2010hx}, we perform $1024$ interpolations between snapshots and finally integrate these to estimate Eq.~(\ref{eq:6}). 

\begin{figure}[t]
  \centering
  \includegraphics[width=0.48\textwidth]{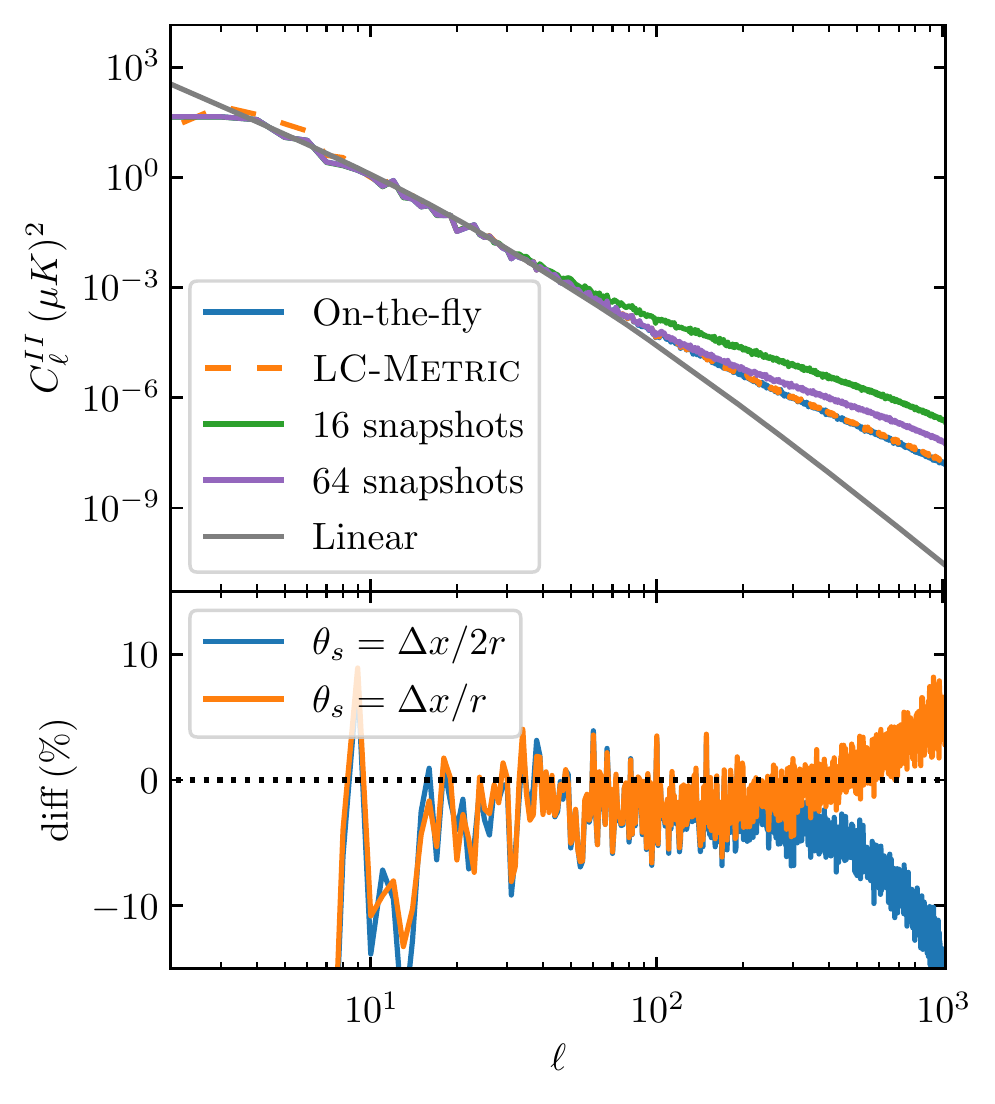}
  \caption{\label{fig:ISW_pow} Upper panel: Power spectrum of the ISW field calculated by integrating snapshots from  Eq.~(\ref{eq:20}) and the reconstructed ``thick'' light cones scheme  using \textsc{LC-Metric} based on 16 snapshots with smoothing angle $\theta_s = \Delta x / 2r$. Lower panel: percent difference between on-the-fly reference and ``thick'' light cones \textsc{LC-Metric} scheme with different smoothing angles $\theta_s$.}
\end{figure}

\begin{figure}[t]
  \centering
  \includegraphics[width=0.48\textwidth]{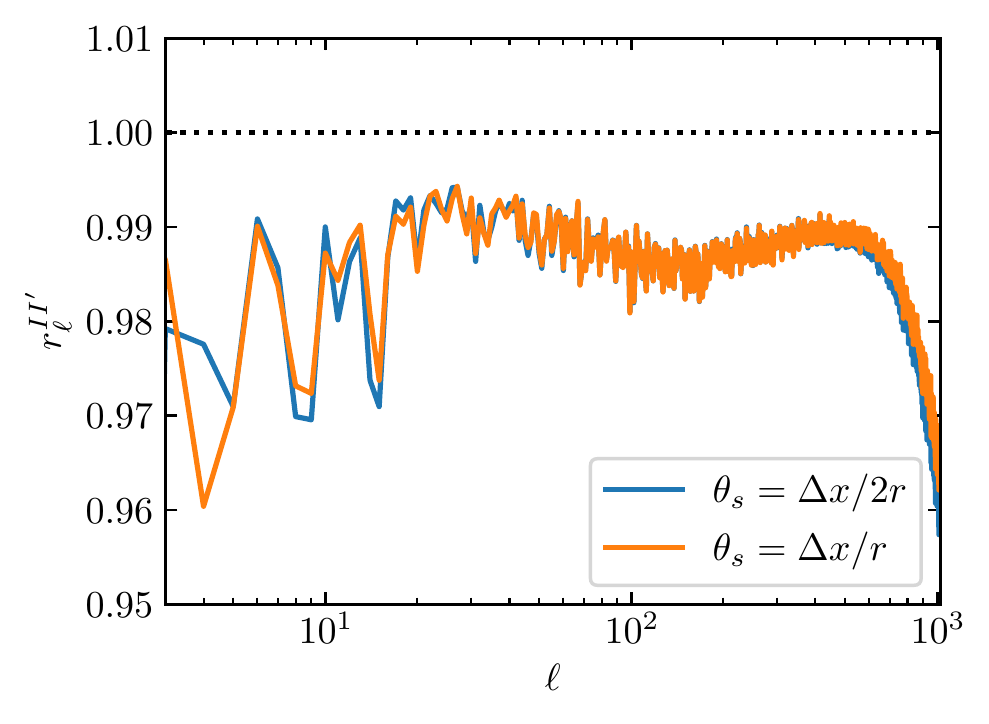}
  \caption{\label{fig:isw_cross} Normalized cross correlation coefficients (Eq.~(\ref{eq:ncc})) of the ISW fields between the on-the-fly scheme based on Eq.~(\ref{eq:20}) and the reconstructing ``thick'' light cones scheme based on \textsc{LC-Metric} with different smoothing angle $\theta_s$. }
\end{figure}

Comparisons of the ISW power spectrum calculated using these different schemes are shown in Fig.~\ref{fig:ISW_pow}.
All of the schemes show agreement with the linear ISW power spectrum at low $\ell$ and begin to deviate from the linear result at $\ell > 60$, consistent with previous studies \cite{Cai:2010hx,Adamek:2019vko,Hassani:2020buk}.
The interpolation scheme based on 16 snapshots significantly overestimates the power compared to the on-the-fly reference spectrum. Increasing the number of snapshots to 64 improves the accuracy but still gives an order of magnitude overestimation.
Even though the ``thick" light cones scheme shown here is based on the light cone constructed from only 16 snapshots, it coincides with the reference spectrum with an accuracy better than $5\%$ percent for $\ell < N_{\rm side}$.
We also note that at the non-linear scale, the ISW spectra highly depends on the manully applied smoothing scale we used as in Eq.~(\ref{eq:24}), whereas the on-the-fly reference scheme includes an intrinsic smoothing scale, whose correspondence to the angular smoothing scale in the \textsc{LC-Metric} is unknown. Instead of looking for the manually applied smoothing angle that best fits the reference on-the-fly scheme, we look for the range of smoothing scale which gives good estimation to the ISW power to the highly non-linear scale. In the lower panel in Fig~\ref{fig:ISW_pow}, we show the ratio between the on-the-fly reference and the \textsc{LC-Metric} scheme under two different angular smoothing scales: $\theta_s = \Delta x / r$ and $\theta_s = \Delta x / 2r$, where the 
later one with a smaller angle naturally produces higher power. Within this range of smoothing scale, an accuracy better than $5\%$ percent for $\ell < N_{\rm side}$ and $15\%$ percent for $\ell < 2N_{\rm side}$ can be achieved. 
On the other hand, as shown in Fig.~\ref{fig:isw_cross}, the normalized cross correlation between the \text{LC-Metric} scheme and the on-on-fly scheme is almost independent of the smoothing scale.
The minor power suppression (lower panel in Fig.~\ref{fig:ISW_pow}) and lack of correlation is again due to the insufficient angular resolution and also the systematics in our construction scheme of light cones from snapshots.
See Appendix \ref{subsec:isw_conv} and \ref{sec:lc_from_snaps} for more details on the light cone construction scheme.

\section{Summary}
\label{sec:summ}

In this work, we have presented a novel scheme \textsc{LC-Metric}, which is able to quantify cosmic observerbles by post-processing simulated large-scale structure data. This reconstruction procedure is based on obtaining solutions to the linearized Einstein's equations on light cone grids with a relaxation scheme; it is also a flexible scheme that is able to process ordinary snapshots or light cone outputs. 

We have also generated mock cosmic observables including the weak-lensing convergence field and the (nonlinear) ISW effect based on the reconstructed metric potential provided by \textsc{LC-Metric}. 
For the weak-lensing calculation, under the Born approximation, the lensing power spectrum calculated explicitly from the reconstructed metric agrees,  typically at a sub-percent level, with the equivalent spectrum calculated via the traditional thin-lens approach, although percent-level disagreement is found on large scales.
Both results are in general agreement with the prediction given by the Halofit prescription over a wide range of scales. For the estimation of the ISW and Rees-Sciama effects, as a post-processing method, the \textsc{LC-Metric}-based ``thick'' light cone scheme is able to model the ISW signal in a highly nonlinear regime and maintain a better than $5\%$ precision for $\ell < N_{\rm side}$ with a proper smoothing angle when compared to a high-precision reference on-the-fly calculation.
When compared to other post-processing schemes such as the one introduced in \cite{Cai:2010hx}, \textsc{LC-Metric} is able to provide a much more precise prediction of nonlinear ISW contributions, which include the Rees-Sciama and moving-lens effects, based on data from considerably fewer N-body simulation snapshots.
Since weak-lensing observables mostly depend on the angular gradient (Laplacian) of the gravitational fields, whereas the ISW effect depends on the time (radial) gradient, being able to predict these two effects precisely demonstrates the utility of the \textsc{LC-Metric} prescription.
Therefore, we conclude that \textsc{LC-Metric} is able to reconstruct the metric accurately in both linear and highly nonlinear regimes, and hence it is suitable for simulating data relevant to current and future large-scale surveys.

There are still several improvements that could be made to the \textsc{LC-Metric} in the future. Firstly, even though our current implementation for the \textsc{LC-Metric} only supports running on a shared memory node, the algorithm can be easily implemented to support distributed memory, which will relieve the relatively stringent memory requirements (currently $\mathcal{O}(N\log N)$, where $N$ is the number of spherical grids).
Secondly, the overall accuracy of the fields, especially the accuracy of $\Xi$ term in Eq.~(\ref{eq:30}), can be potentially improved by a more accurate deposition technique of the radial velocity field, e.g., Delaunay Tessellation Field Interpolation (DTFE) \cite{astro-ph/0011007,1105.0370}.
Finally, a general-relativistic ray-tracing component has been implemented, which is necessary in order to quantify higher-order and post-Born effects. We will investigate those effects in a future study.

Upcoming large-area and high-precision galaxy and CMB surveys will deepen our understanding of the Universe as well as bring unprecedented challenges for modeling general-relativistic cosmic observables such as weak-lensing, the Rees-Sciama effect, moving-lens effect, kSZ effect, etc. The \textsc{LC-Metric} scheme provides a universal way to model these effects while maintaining a high level of precision even in the highly nonlinear regime, making it well-suited to testing and analyzing catalogs for these surveys in the future. Knowledge of the spacetime metric will also facilitate the estimation of other relativistic or lightcone projections effects that have not been properly quantified and measured, as listed in \cite{1105.5280}. Furthermore, the post-processing-based feature of \textsc{LC-Metric} will allow for consistent analysis of N-body simulations or emulations running with different physics.
We expect that this new scheme will provide us with some insight into reconstructing metric information from observations while simultaneously accounting for different relativistic and projection effects. The reconstructed cosmic density and velocity fields are needed for such reconstruction; while the former may be estimated from galaxy or halo catalogues \cite{Wang:2008wx,Munoz-Cuartas:2011gik}, the later may be restored from e.g. kSZ \cite{1707.08129} or observations of the moving lens effect.

Finally, we note a recent study \cite{2107.13008} that points out the value of determining the ``boosted potential" in order to improve our understanding of several aspects of structure formation. This scheme relies on quantifying the gravitational potential in a boosted frame. Our new approach for constructing the gravitational potential on the light cone is suitable for extracting the boosted potential based on light cone data. We reserve further investigation in this direction for future work.

To summarize, we have introduced the \textsc{LC-Metric} scheme for constructing the spacetime metric from general N-body-code outputs.
We have demonstrated its accuracy  by comparing it to conventional methods in quantifying the weak-lensing convergence and the nonlinear-ISW effects.  Therefore, it is possible to incorporate \textsc{LC-Metric} with general N-body catalogs and facilitate estimation of higher-order and post-Born relativistic effects on cosmological observables.

\section*{Acknowledgements}

We thank Stefano Anselmi and Selim Hotinli for useful  discussions and feedback. JBM is partly supported by a Department of Energy grant DE-SC0017987 to the particle astrophysics theory group at WUSTL.
GDS is partly supported by a Department of Energy grant DE-SC0009946 to the particle astrophysics theory group at CWRU.
The simulations and analyses in this work used the Data Storage Platform and Scientific Compute Platform operated by the Research Infrastructure Services (RIS) of Washington University in St. Louis. 

\bibliography{inspire_auto_references,manual_references}

\appendix

\section{Numerical validations}
\label{sec:conv}

In this appendix, we will show several convergence tests we have performed to validate our results.

\subsection{Convergence of the multi-grid solver}
\label{subsec:solver}

\begin{figure}[ht]
  \centering
  \includegraphics[width=0.48\textwidth]{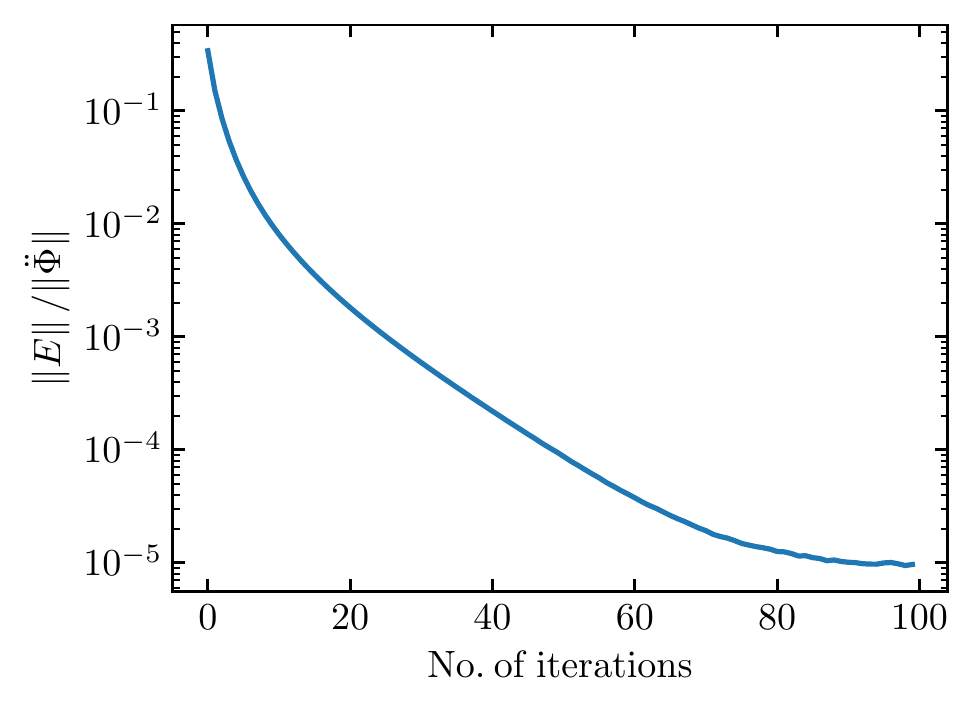}
  \caption{\label{fig:E_viol} Evolution of L2 norm of relative error of Eq.~(\ref{eq:28}) with respect to the number of V-cycles performed.}
\end{figure}

We estimate the convergence of our \textsc{LC-Metric} solver from the L2 norm of the relative error $\left\Vert E\right\Vert / \Vert \ddot{\Phi}\Vert$ and show the result in Fig.~\ref{fig:E_viol}, where $E$ is the error of Eq.~(\ref{eq:28}) and $\Vert \ddot{\Phi}\Vert$ represents the L2 norm of its left hand side. The radial and angular resolution of the light cone grids are $N_{\eta} = 1024$ and $N_{\rm side} = 512$ respectively, and the number of levels of the multi-grid hierarchy is 5. Our multi-grid scheme will reduce the relative error very efficiently until reaching a plateau. The final plateau in Fig.~\ref{fig:E_viol} is limited by the machine precision of 32-bit float used in the solver. The total memory consumption is about $10$ times the memory consumption of the light cone grids, and the time consumption for 100 iterations is about $20$ hours on a computer node with 28 CPUs. 
\subsection{Numerical convergence of the interpolations to HEALPix grids}
\label{subsec:interp_conv}

\begin{figure}[ht]
  \centering
  \includegraphics[width=0.48\textwidth]{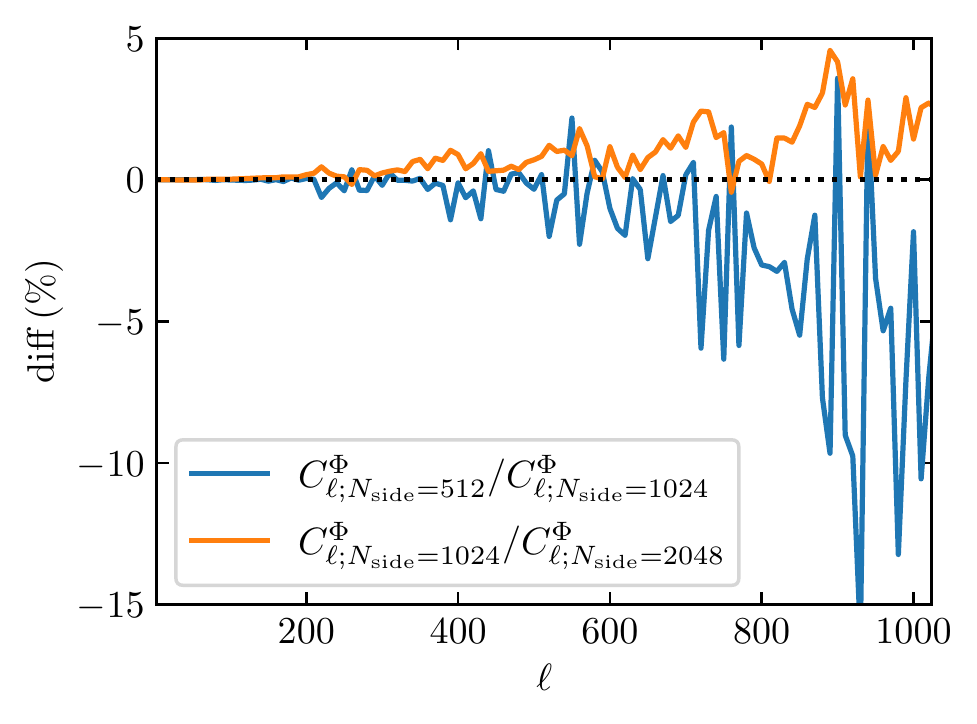}
  \caption{\label{fig:interp_conv} The ratio of the angular power spectrum of $\Phi$ interpolated from Cartesian grids with different HEALPix resolutions $N_{\rm side}$. The radius for such interpolations is set according to $r = \sqrt{(A / \Omega)}$, where $A$ is the Cartesian resolution square $\mathrm{d}x^2$ and $\Omega$ is the HEALPix solid angle for $N_{\rm side} = 512$.}.
\end{figure}

When calculating a reference power spectrum of the gravitational potential  or ISW (see Section \ref{subsec:comp} and \ref{subsec:isw}), or when setting initial conditions for the solver, interpolations of fields from Cartesian grids to HEALPix grids are performed. 
We test the accuracy of such operations by linearly interpolating the same potential fields on Cartesian grids with different angular resolutions.
As shown in Fig. \ref{fig:interp_conv}, by comparing the ratio of angular power between different $N_{\rm side}$s, we can estimate the accuracy of such interpolations from Cartesian grids to HEALPix grids, which is better than $5\%$ for $\ell < N_{\rm side}$, and better than $20\%$ for $\ell < 2 N_{\rm side}$. This limits the numerical precision of both reference power spectra and the $\textsc{LC-Metric}$ scheme.

\subsection{Numerical convergence of the gravitational potential}
\label{subsec:phi_conv}

\begin{figure}[ht]
  \centering
  \includegraphics[width=0.48\textwidth]{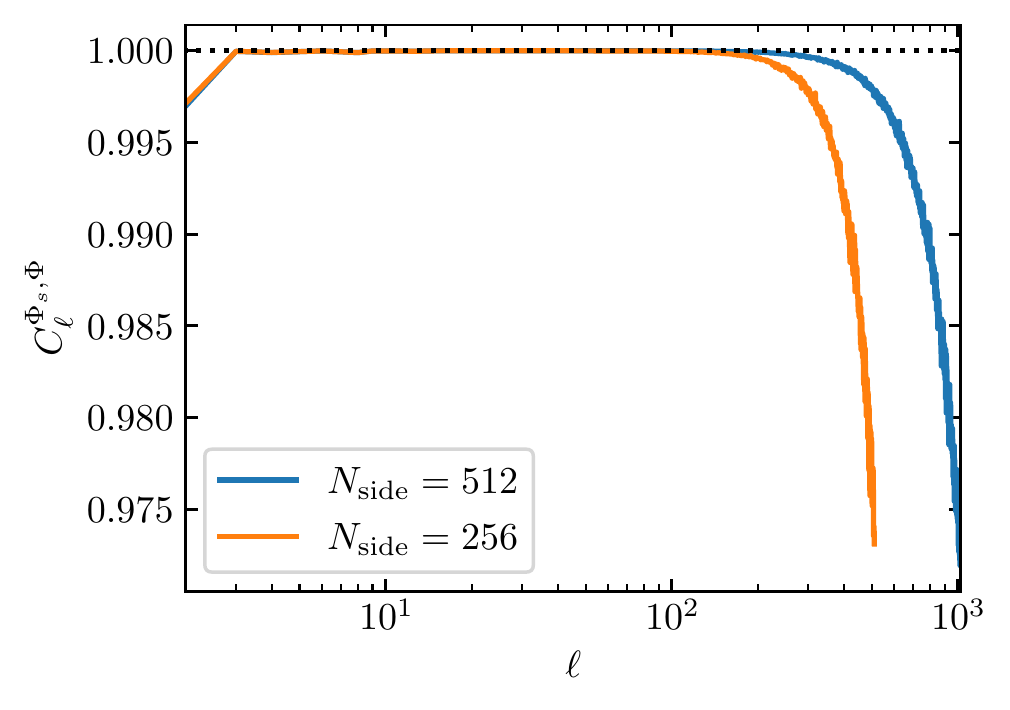}
  \caption{\label{fig:phi_conv} The same as Fig. \ref{fig:phi_cross}. Normalized cross correlation coefficients between the reconstructed $\Phi_{\rm LC}$ and the $\Phi_{\rm snap}$ under different $N_{\rm side}$. }.
\end{figure}

Even though the precision of power spectra at higher $\ell$ is primarily limited by linear interpolation accuracy as described in the previous subsection, as suggested by Fig. \ref{fig:phi_conv}, improving the angular resolution $N_{\rm side}$ can potentially improve the precision at higher $\ell$, since the cross correlation at the same $\ell$ is significantly larger when comparing the result of $N_{\rm side} = 512$ with $N_{\rm side} = 256$. 

\subsection{Numerical convergence of weak-lensing}

\begin{figure}[ht]
  \centering
  \includegraphics[width=0.48\textwidth]{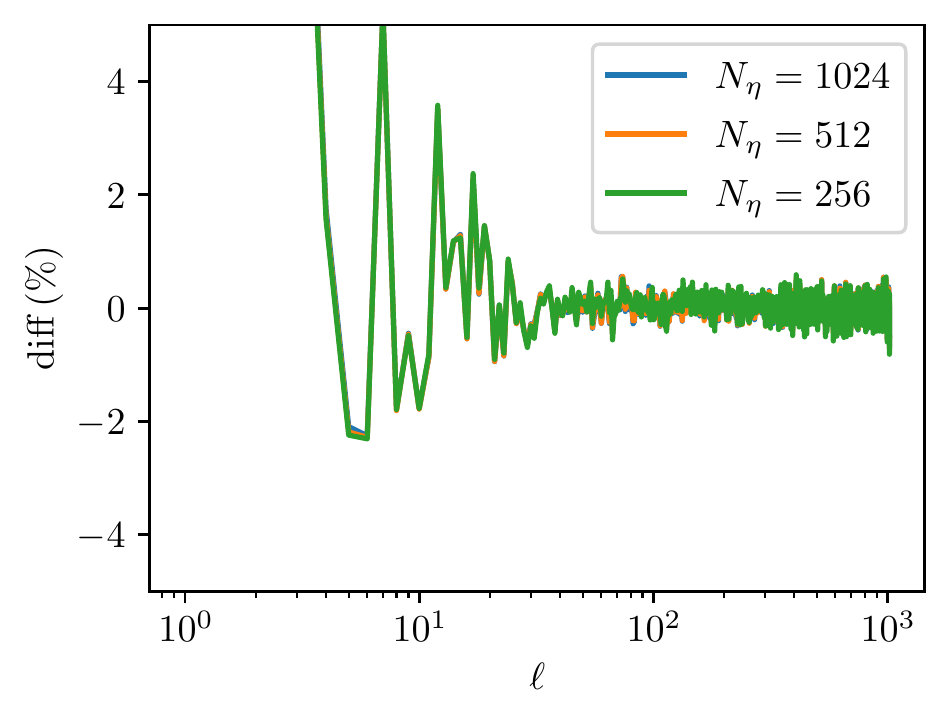}
  \caption{\label{fig:lensing_conv} Ratio of the thin-lens approach to the scheme based on the \textsc{LC-Metric} reconstruction with different radial resolution $N_{\eta}$, same as lower panel in Fig.~\ref{fig:lensing_pow}}.
\end{figure}

The weak-lensing observables extracted from reconstructed metric from light cone data are independent of the radial resolution chosen for our coordinate system.
The test of the convergence against different radial resolutions is shown in Fig.~\ref{fig:lensing_conv}. No noticeable differences are identified for reasonable resolutions from $256$ to $1024$, suggesting the robustness of the numerical value of weak-lensing convergence calculated by the \textsc{LC-Metric} scheme.

\subsection{Numerical convergence of the ISW and Rees-Sciama effects}
\label{subsec:isw_conv}

\begin{figure}[ht]
  \centering
  \includegraphics[width=0.48\textwidth]{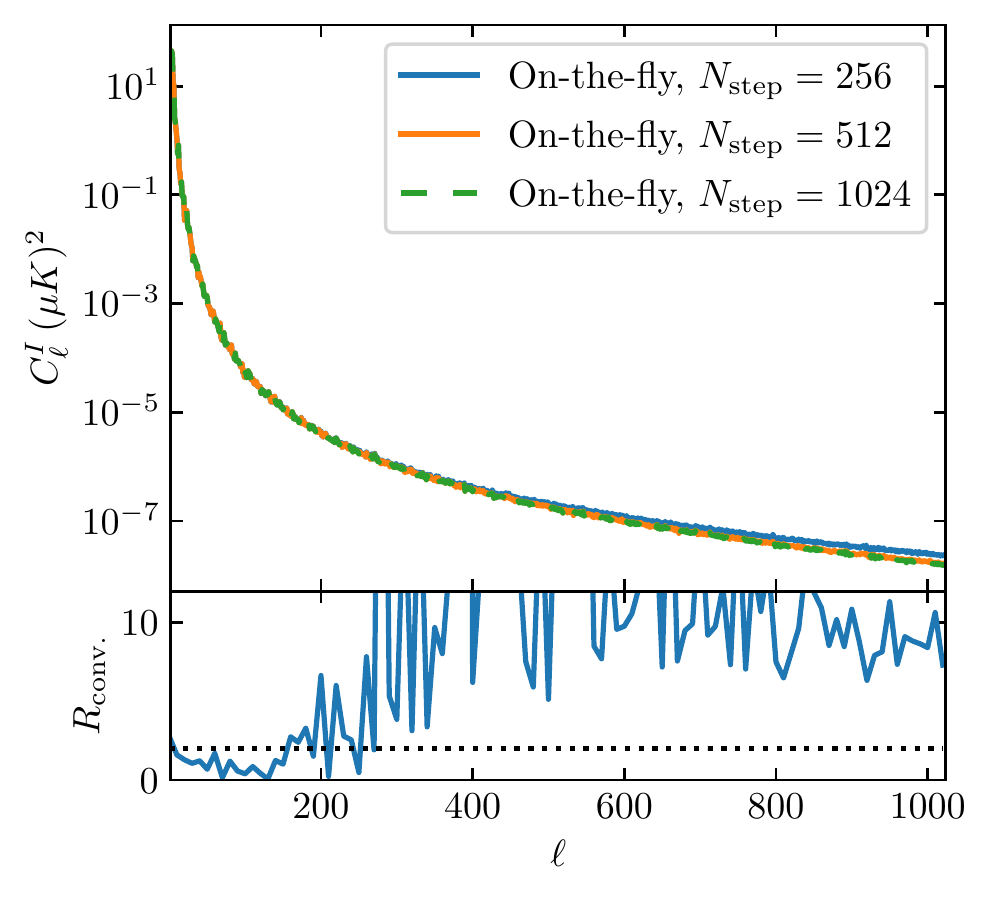}
  \caption{\label{fig:isw_ref_conv} Upper panel: ISW and Rees-Sciama effects calculated through on-the-fly method against number of time steps. Convergent result is achieved when $N_{\rm steps} = 1024$. Lower panel: Convergence rate $R_{\rm conv.}$ defined in Eq.~\ref{eq:12}, the result is higher than the linear convergence rate (dotted line) for $\ell > 200$}.
\end{figure}

The reference ISW power spectrum calculated through the on-the-fly scheme in Fig.~\ref{fig:ISW_pow} is extracted from a N-body simulation performing $N_{\rm steps} = 1024$ time steps. Fig.~\ref{fig:isw_ref_conv} validates this choice of number of time steps by comparing the ISW power with different number of time steps since the results of $N_{\rm steps} = 512$ and $N_{\rm steps} = 1024$ overlap. We also calculate the convergence rate $R_{\rm conv}$, defined as
\begin{equation}
\label{eq:12}
R_{\rm conv} =\frac{\left|f_{N_{m}} - f_{N_{c}}\right|}{\left|f_{N_{c}} - f_{N_{f}}\right|},
\end{equation}
where $f_{N_{c}}$, $f_{N_{m}}$, $f_{N_{f}}$ are values calculated at resolutions $N_{c}$, $N_{m}$, $N_{f}$, which are from coarsest to finest, and shown in the lower panel in Fig.~\ref{fig:isw_ref_conv} that the rate is beyond the linear convergence for the $\ell > 200$ region.

\begin{figure}[ht]
  \centering
  \includegraphics[width=0.48\textwidth]{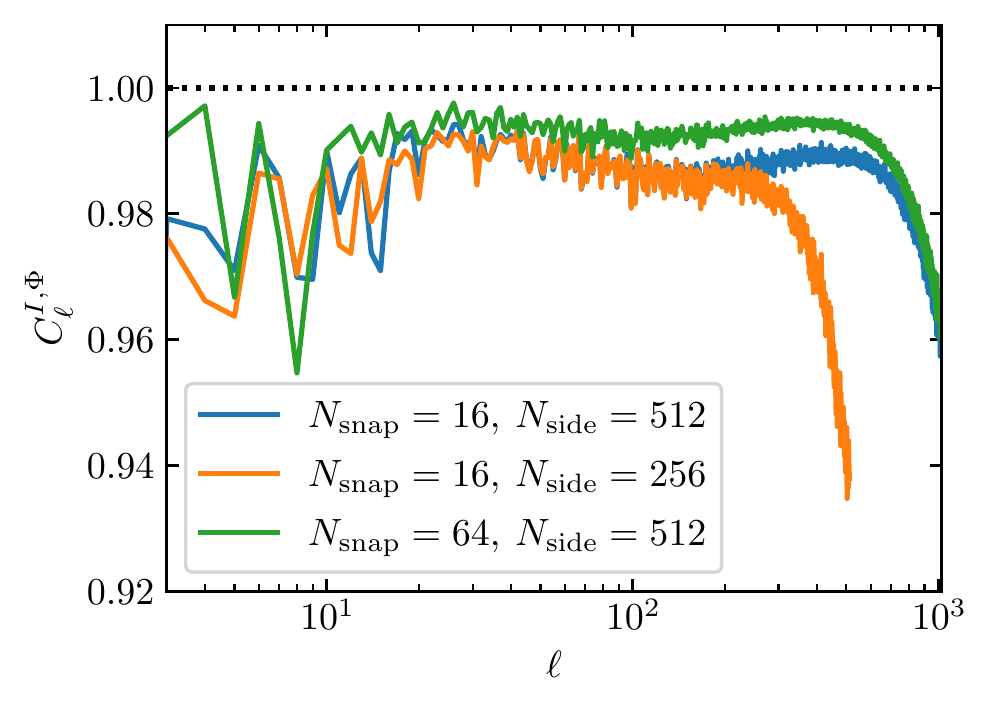}
  \caption{\label{fig:isw_conv} Same as Fig.~\ref{fig:isw_cross}. Variations of normalized cross correlation coefficients of the ISW fields between the snapshots scheme and  the  reconstructing  ``thick''  light cones against different numbers of snapshots and different angular resolutions $N_{\rm side}$.}
\end{figure}

The minor lack of correlation between the reference ISW power and the reconstructed one through ``thick'' light cones (Fig.~\ref{fig:isw_cross}) is likely due to the systematic bias when creating light cone data from snapshots (see Appendix~\ref{sec:lc_from_snaps} for more detail). This can be confirmed by comparing the blue and green curve in Fig.~\ref{fig:isw_conv}, where employing more snapshots to construct ``thick'' light cones will improve the correlation. The suppression of power at higher $\ell$ is again due to the limited precision of the linear interpolations, which can be improved by employing a higher angular resolution according to Fig.~\ref{fig:isw_conv}.

\section{light cone construction from snapshots}
\label{sec:lc_from_snaps}

The scheme integrated in the \textsc{LC-Metric} to construct light cone data from snapshots is similar to the particle extrapolations scheme mentioned in \cite{Hollowed:2019jjn}. Several improvements have been implemented to enhance the construction accuracy. Here, we will give our full algorithm on constructing light cone from snapshots.

For each particle inside the light cone, which has comoving position $\bm{r}_i$, peculiar velocity $\bm{v}_i$ and acceleration $\bm{a}_{i}$ on the snapshot at conformal time $\tau_i$, the time it crosses the light cone is $\tau + \delta\tau$, and the corresponding comoving radii after the time $\delta \tau$ is
\begin{eqnarray}
\label{eq:18}
  |\bm{r}| &=& \left|\bm{r}_i + \bm{v}_i \delta \tau + \frac{\bm{a}_{i}}{2} \tau^2\right| \nonumber \\
      &\approx& \left| \bm{r}_i\right| + \delta \tau\frac{\bm{r}_{i}\cdot \bm{v}_i}{\left|\bm{r}_{i}\right|} \nonumber \\
      &&- \delta\tau^{2} \left[\frac{(\bm{r}_{i}\cdot \bm{v}_{i})^{2}}{2 \left|\bm{r}_{i}\right|^{3}}
                + \frac{\bm{v}_{i}\cdot \bm{v}_{i}}{2 \left|\bm{r}_{i}\right|} + \frac{\bm{a}_{i}\cdot \bm{r}_{i}}{2 \left|\bm{r}_{i}\right|}\right],
\end{eqnarray}
note we have kept up-to 2nd-order terms, and the acceleration $\bm{a}_i$ can be estimated from difference between peculiar velocities between two adjacent snapshots from
\begin{equation}
\label{eq:8}
\bm{a}_{i} = (\bm{v}_{i+1} - \bm{v}_{i}) / \Delta\tau,
\end{equation}
where $\Delta\tau$ is the difference in the conformal time $\tau$ between these two snapshots.
Additionally, since the background light cone satisfies $\tau = -|\bm{r}| $, we can write down the following relation
\begin{equation}
\label{eq:5}
|\bm{r}| = \tau_i -\delta \tau.
\end{equation}
After plugging-in Eq.~\eqref{eq:18} to Eq.~\eqref{eq:5}, we can solve $\delta \tau$ from the quadratic equation. However, the following optimizations are necessary to construct the light cone accurately:

\begin{itemize}
\item For particle $p_j$ outside the light cone on a snapshot $s_i$, we do not extrapolate this particle since it will never enter the light cone in the future.
  
\item For particle $p_j$ inside the light cone on a snapshot $s_i$, if $p_j$ inside the next snapshot (snapshot $s_{j+1}$) is still inside the light cone, we do not extrapolate this particle.
\item We use the 2nd-order quadratic approximation to solve Eq.~(\ref{eq:18}) instead of the exact solution to avoid possible numerical instabilities.
\item If the solution of $\delta\tau$ satisfies $\delta\tau > \Delta \tau$, no extrapolation will be performed.
\end{itemize}

Note that this scheme will tend to omit some particles crossing the light cone because of the limited accuracy of the quadratic interpolation, and therefore will introduce some systematic bias to the reconstructed light cone. Investigating an improved scheme to reduce this bias is left for future study.

\end{document}